\documentclass[twocolumn,groupedaddress,superscriptaddress,amsmath,amssymb,floatfix,aps,prl,showpacs]{revtex4}
\usepackage{amsmath}    
\usepackage{graphicx}   
\usepackage{verbatim}   
\usepackage{color}      
\usepackage{subfigure}  
\usepackage{hyphenat}

\begin{document}
\bibliographystyle{prsty} 

\title{How biochemical resources determine fundamental limits in cellular sensing}
\author{Christopher C. Govern}
\author{Pieter Rein ten Wolde}
\affiliation{FOM Institute AMOLF, Science Park 104, 1098 XG Amsterdam, The Netherlands.}

\begin{abstract}

Living cells deploy many resources to sense their
  environments, including receptors, downstream signaling molecules,
  time and fuel. However, it is not known which resources
  fundamentally limit the precision of sensing, like weak links in a
  chain, and which can compensate each other, leading to trade-offs
  between them. We show by modeling that in equilibrium systems the
  precision  is limited by the number of
  receptors; the downstream network can never increase precision. This
  limit arises from a trade-off between the removal of extrinsic noise
  in the receptor and intrinsic noise in the
  downstream network. Non-equilibrium systems can lift this trade-off
  by storing the receptor state over time in chemical modification
  states of downstream molecules. As we quantify for a push-pull
  network, this requires i) time and receptors; ii)  downstream
  molecules; iii) energy (fuel turnover) to drive
  modification. These three resource classes cannot compensate each
  other, and it is the limiting class which sets the fundamental
  sensing limit. Within each class, trade-offs are possible. Energy
  allows a power-speed trade-off, while time can be traded against
  receptors.
\end{abstract}

\maketitle

Biochemical networks are the information processing devices of life.
Like any device, they require resources to be built and
run. Components are needed to construct the network, space is required
to accommodate the components, time is needed to process the
information, and energy is required to make the components and operate
the network. These resources constrain the design and performance of any biochemical
network. Yet, it is not clear which resources are indispensable, thus
fundamentally limiting the performance of the network, and which
resources might trade-off against each other.  Here we consider the
interplay  among cellular resources, network design, and performance in a canonical biochemical function,
namely sensing the environment.  

Living cells can measure chemical concentrations with extraordinary
precision (\citealp{berg1977,Sourjik:2002fk,Ueda:2007uq}), raising the
question what sets the fundamental limit to the accuracy of chemical
sensing (\citealp{berg1977}). Cells measure chemical concentrations
via receptors on their surface.  These measurements are inevitably
corrupted by noise that arises from the stochastic arrival of ligand
molecules by diffusion and from the stochastic binding of the ligand
to the receptor.  Berg and Purcell pointed out that the sensing error
is fundamentally bounded by this noise extrinsic to the cell, but that
cells can reduce the error by taking multiple independent
measurements, mitigating the risk that any one is corrupted by a noisy
fluctuation (\citealp{berg1977}). One way to increase the number of
measurements is to add more receptors to the surface
(\citealp{berg1977,levinepre2007}). Another is to take more
measurements per receptor over time; in this approach, the cell infers
the concentration not from the instantaneous number of ligand-bound
receptors but rather from the time-average receptor occupancy over an
integration time $T$ (\citealp{berg1977,bialek2005, levinepre2007,
  levineprl2008,wingreen2009,levineprl2010,mora2010,Govern2012}).

This time integration has to be performed by the signaling networks
that transmit the information from the surface of the cell to its
interior (\citealp{Govern2012}). To
reach the fundamental limit on the accuracy of sensing, these networks
have to remove the extrinsic noise in the receptor state as much as
possible. Signaling networks, however, are also stochastic in nature,
which means that they will also add noise to the transmitted signal.
Most studies on the accuracy of sensing have ignored this intrinsic
noise of the signaling network. They essentially assume that the
intrinsic noise can be made arbitrarily small and that the extrinsic
noise in the input signal can be filtered with arbitrary precision by
simply integrating the receptor signal for longer.  Yet, the extrinsic
and intrinsic noise are not generally independent
(\citealp{TanaseNicola2006}).  Indeed, what resources are required to
simultaneously remove the extrinsic and intrinsic noise is not
understood.

While the work of Berg and Purcell and subsequent studies identify
time and the number of receptors as resources that limit the accuracy
of sensing, the fundamental limits that have emerged ignore the cost
of making and operating the signaling network.  Making proteins is
costly; producing proteins that confer no benefit to the cell has been
shown to slow down bacterial growth (\citealp{Dekel2005}).  They also
take up valuable space that might be used for other important
processes, either on the membrane or inside the cytoplasm. Both are
highly crowded, with proteins occupying $25-75\%$ of the membrane area
(\citealp{Linden2012}) and $20-30\%$ of the cytoplasmic volume
(\citealp{Ellis2001}). Moreover, many signaling networks must be
driven out of thermodynamic equilibrium by the continuous turnover of
fuel molecules such as ATP, leading to the dissipation of heat. Fuel
is essential for network functions such as bistability, oscillations, and
kinetic proofreading
(\citealp{Hopfield:1974uo,Ninio:1975vv,Qian:2006dl}), and can be
important for adaptation (\citealp{Tu2012, Endres}). However, whether
there exists a fundamental relationship between energy and sensing,
independent of the design of the signaling network inside the cell,
remains unclear (\citealp{Mehta2012,Barato2013,Skoge:2013fq}).

In this manuscript we derive how the accuracy of sensing depends
  on not only time and the number of receptors, but also on the
  resources required to build and operate the downstream signaling
  network: the copies of signaling molecules and fuel.  This allows
us to address the following questions: How do the sensing limits set
by the latter resources compare to the canonical limit of Berg and
Purcell, which is set by the resources time and the number of
receptors?  How does the limit set by one resource depend on the
levels of the other resources?  Can resources compensate each other to
achieve a desired sensing precision, leading to trade-offs between
them, or are the limits set by the respective resources fundamental,
{\it i.e.}  independent of the levels of the other resources? And how
do the limits depend on the design of the signaling network?  The
  relationship between the accuracy of sensing, the design of the
  network, and the resources required to build and operate it---time,
  energy and protein copies---underlies the design principles of
  biochemical sensing systems.

We first study the relationship between sensing precision, network
design, and resources, for systems that are not driven out of
thermodynamic equilibrium, consuming no fuel.  We find that
these equilibrium networks can time-integrate the receptor signal to remove
the extrinsic noise in it, analogous to the mechanism described by
Berg and Purcell. Clearly, fuel is not a fundamental resource for
sensing or removing extrinsic noise.  However, using the
fluctuation-dissipation theorem, we will show that equilibrium networks
face a fundamental trade-off between the removal of extrinsic noise in
the receptor state and the suppression of intrinsic noise in the
processing network: decreasing one source of noise
  necessarily increases the other.  As a result, the accuracy of
sensing is fundamentally limited by the number of receptors; in
equilibrium networks, adding downstream components can never improve
the sensing precision.

To improve the sensing accuracy beyond the limit set by the number of
receptors, it is essential to break the trade-off between extrinsic
and intrinsic noise. As we show, this requires a fundamentally
different sensing mechanism. Instead of using the receptors to harvest
the energy of ligand binding, as in the equilibrium sensing mode, the
receptors should be used as catalysts to modify downstream read-out
molecules. This non-equilibrium strategy, however, uses not only
receptors but requires also time, copies of downstream read-out
molecules, and fuel turnover.

We quantify the limits that arise from each of the resources---copies
of receptors and downstream molecules, time, and fuel---for a
canonical signaling motif, a receptor that drives a Goldbeter-Koshland
push-pull network (\citealp{Goldbeter:1981qf}).  Push-pull networks
are ubiquitous in prokaryotic and eukaryotic cell signaling
(\citealp{Alon:2007tz}): examples include GTPase cycles, as in the Ras
system (\citealp{Pylayeva-Gupta:2011kx}), and phosphorylation cycles,
as in mitogen-activated-protein-kinase (MAPK) cascades
(\citealp{Chang:2001uq}) or in two-component systems like the
chemotaxis system of {\em Escherichia coli} (\citealp{Stock:2000ve}).
We find that the resource limitations of these systems emerge
naturally when the signaling networks are viewed as devices that
discretely, rather than continuously, sample the receptor state via
collisions of the signaling molecules with the receptor proteins.
This analysis reveals that three classes of resources are required: i)
time and receptors; ii) copies of downstream molecules; and iii)
fuel. Indeed, these classes cannot compensate each other: each imposes
a sensing limit, and it is the limiting class that imposes the
fundamental limit on the accuracy of sensing.  However, there can be
trade-offs within each class of resources.  Receptors and time trade
off against each other in achieving a desired sensing accuracy and
power and response time trade off against each other to meet the
energy requirement for taking a measurement.  We end by discussing how
our findings apply to specific signaling systems and how our results
on push-pull networks generalize to more complex networks involving
cascades, and positive and negative feedback. In particular, our
analysis  leads to a concrete prediction for the optimal resource
allocation, which we test for the {\em E. coli} chemotaxis system.

\section{Results}

Consider a cell with $R_T$ receptors on its surface that independently
bind ligand, $R + L \rightleftharpoons RL$.  The cell senses the
ligand concentration $c$ based on the instantaneous level of a
downstream read-out molecule $x$ at some time.  Via error propagation, the cell's uncertainty
about $c$ is then (\citealp{bialek2005}):
\begin{equation}
\label{eq:error}
\left( \frac{\delta c}{c} \right)^2 = \frac{1}{c^2} \frac{\sigma_{x}^2}{ \left( \frac{d\overline{x}}{dc} \right)^2} = \frac{\sigma_{x}^2}{ \left( \frac{d \overline{x}}{d\mu_L} \right)^2 },
\end{equation}
where $\mu_L = \mu_0 + \log c$ is the ligand's chemical potential.
The uncertainty is low if the average read-out level $\bar{x}$
responds sensitively to changes in ligand concentration, as
measured by the gain $d \bar{x}/d c$, but is not noisy, as measured by
the variance $\sigma^2_x$.

If the receptor-ligand complex itself is taken as the read-out, then the error is:
\begin{equation}
\label{eq:receperr}
\left( \frac{\delta c}{c} \right)^2 =  \frac{1}{p(1-p)} \frac{1}{N_I}
= \frac{1}{p(1-p)} \frac{1}{R_T}, 
\end{equation}
 since $\sigma_{RL}^2 = \frac{d \overline{RL}}{d\mu_L} = R_T p (1-p)$,
where $p$ is the probability a receptor is bound to ligand. Indeed,
$1/(p(1-p))$ is the ``instantaneous error'', {\it i.e.}  the
  sensing error based on a single concentration estimate via a
  single receptor. Because each receptor provides an independent
concentration  measurement (\citealp{levinepre2007}), the total number of independent
measurements is $N_I=R_T$. Clearly, the sensing error is limited by
the total number of receptors on the membrane.

\section{Trade-offs in equilibrium sensing}

Cells can reduce the error in Eq. \ref{eq:receperr} with downstream
networks that time-integrate over the history of receptor states
(\citealp{berg1977}).  Key to the ability of networks to
time-integrate is a memory of these past states, implemented, for
example, by a long-lived molecular species or a signaling cascade that
delays the signal (\citealp{Govern2012}).  Equilibrium systems can
have these and hence have memory of the past receptor states.  Thus,
we might expect that equilibrium networks can reduce the sensing error
past the bound set by the number of receptors at the expense of
downstream signaling molecules.

We consider cytoplasmic read-out molecules $x$ that bind ligand-free
receptors: $R + L \rightleftharpoons RL$, $R + x \overset{k_{\rm f}}
{\underset{k_{\rm r}}\rightleftharpoons} Rx$.  Solving the associated
Langevin equations (\emph{Materials and Methods}) shows that the dynamics of the
output around its mean $\overline{x}$ is given by the time-integrated fluctuations
$\delta RL(t)$ in the receptor state plus noise $\eta(t)$ due to the receptor-read-out binding:
\begin{equation}
\label{eq:lang}
\delta x(t) = \int_{-\infty}^t dt' e^{-(t-t')/\tau_{I}} \left[\beta\delta RL(t')  + \eta(t') \right],
\end{equation}
where $\beta = k_{\rm f} \overline{x}$ and  $\tau_{I} = \frac{1}{k_{\rm f}
  \left( \bar{x} + \bar{R} \right) + k_{\rm r}}$ is the integration time.  The latter can be made arbitrarily large by slowing down the read-out
dynamics, i.e. by lowering $k_{\rm f}$ and $k_{\rm r}$. This suggests
that equilibrium networks can completely filter the extrinsic noise in
the receptor states  and reduce the sensing error to zero. However,
the idea that the sensing error can be reduced to zero ignores the fact that in these equilibrium systems
ligand-receptor binding and receptor-read-out binding are
coupled. In this specific system, these reactions are coupled
  because the
read-out and the ligand compete for binding to the receptor.

To elucidate how the coupling between receptor-ligand binding and
receptor-read-out binding compromises sensing in equilibrium
networks, we determine the total sensing error.  From
Eq. \ref{eq:lang}, the variance of the output $\sigma^2_x= \langle \delta x \rangle^2$ can be
written as the sum of the extrinsic noise $\sigma_{\rm ex}^2 \equiv \beta^2
K_{\delta RL,\delta RL}$ and the intrinsic noise $\sigma_{\rm in}^2 \equiv
\beta K_{\delta RL,\eta}+ K_{\eta ,\eta}$, where $K_{A,B} =
\int_{-\infty}^{t} \int_{-\infty}^{t} e^{-(t-t_1')/\tau_{\rm I}}
C_{A,B}(t_1',t_2') e^{-(t-t_2')/\tau_{\rm I}} dt_1 dt_2$ with the correlation function
$C_{AB}(t_1,t_2) = \langle A(t_1) B(t_2) \rangle$. Combining
$\sigma^2_x$ with the gain $d\overline{x}/dc$ gives the sensing error
for this network (Eq. \ref{eq:error}).  Analytically minimizing the
result, we find that it is never lower than the bound set by the
number of receptors (i.e. $N_I \le R_T$ in Eq. \ref{eq:receperr}),
regardless of the integration time or other parameters of the network (\emph{SI Text}).  This raises the paradox of a network that time-integrates
the receptor fluctuations yet cannot reduce the sensing error with it.
The resolution of the paradox is that in equilibrium systems the
intrinsic and extrinsic noise are not independent, precisely because
receptor-ligand and receptor-readout binding are coupled. As a result,
the fluctuations in the receptor state and the read-out become
correlated; $C_{\delta RL,\eta}(t_1,t_2)$ is not zero
(\citealp{TanaseNicola2006}). Because of these correlations, 
  equilibrium networks face a fundamental trade-off between
  the removal of extrinsic noise in the receptor state and the
  suppression of intrinsic in the downstream signaling network. In an
  optimally designed network that minimizes the sensor error,
  increasing the integration time reduces the extrinsic noise, but
  also increases the intrinsic noise by at least the same amount.

Signaling networks are usually far more complicated than a single
read-out molecule that binds the receptor, and it has been shown that
additional network layers can reduce the sensing error
(\citealp{Govern2012}).  This raises the question whether a more complicated
equilibrium network can overcome the limit set by the number of
receptors.  Searching over all possible network topologies to
systematically address this question is difficult, if not impossible.
However, equilibrium systems are fundamentally bounded by the laws of
equilibrium thermodynamics, regardless of their topology. One such law is the
fluctuation-dissipation theorem. Just as a decrease in the viscosity
of a  fluid increases both the noise in a particle's Brownian motion and the sensitivity of its response
to an applied force, so too do modifications in equilibrium
networks affect both the noise in the read-out and
 the sensitivity of its response to changes in the ligand
concentration, {\it i.e.} the gain.

Specifically, for any read-out $x$ in an equilibrium system, the
fluctuation-dissipation theorem implies that the gain $\frac{d
  \bar{x}}{d\mu_L}$ is equal to the covariance $\sigma^2_{x,RL}$ of the fluctuations in the
read-out and the ligand-bound receptor:  $\frac{d
  \bar{x}}{d\mu_L}= \sigma^2_{x,RL}$  (\citealp{Risken1984}).  Then, the sensing error from
any read-out is (Eq. \ref{eq:error}): $\left( \frac{\delta c}{c}
\right)_{x}^2 = \frac{\sigma_x^2}{\left( d\bar{x}/d\mu_L \right)^2} =
\frac{ \sigma_{x}^2 }{ \left( \sigma_{x, RL}^2 \right)^2}$.  If the
receptors themselves are taken as the read-out, the sensing error is
$\left( \frac{\delta c}{c} \right)_{RL}^2 = \frac{1 }{ \sigma_{RL}^2
}$.  By combining these expressions, it follows that no read-out is better for
sensing than the receptors:
\begin{equation}
\left( \frac{\delta c}{c} \right)_{x}^2 =  \frac{ \sigma_{x}^2 \sigma_{RL}^2  }{  \left( \sigma_{x, RL}^2 \right)^2  } \left( \frac{\delta c}{c} \right)_{RL}^2 \ge \left( \frac{\delta c}{c} \right)_{RL}^2
\end{equation}
since the correlation coefficient $ |\rho_{x,RL}| =
|\sigma_{x,RL}^2|/\sqrt{\sigma_x^2 \sigma_{RL}^2} \leq 1$. 

\begin{figure}[b]
\centering
\includegraphics[width=8.5cm]{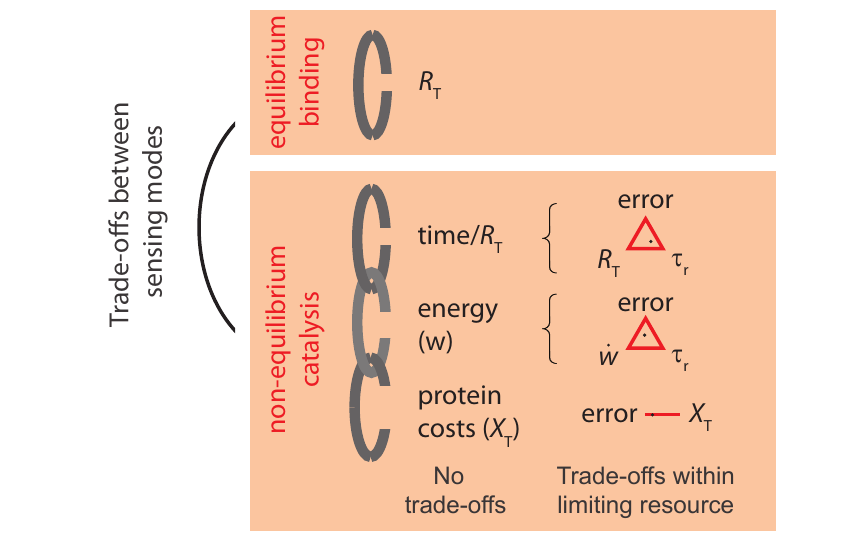}
\caption{ The relationship among network design,
resources, and the precision of biochemical sensing. Cells can use two
distinct modes of sensing and signaling, the equilibrium mode based on
protein binding and sequestration, and the non-equilibrium mode based
on protein modification driven by fuel turnover.  Upper box: In
equilibrium sensing, the sensing accuracy is fundamentally limited by
the number of receptors, regardless of the design of the downstream
newtork. Lower box: In non-equilibrium sensing, the sensing precision
is fundamentally limited by time and receptor copies, energy, and
copies of downstream readouts. These three classes of resources cannot
compensate each other, and it is the limiting resource that sets the
fundamental limit to the precision of sensing. Within each class,
however, trade-offs are possible: Power can be traded against speed to
meet the energy requirement for reaching a desired sensing accuracy,
while time can be traded against the number of receptors.\label{fig:eqneqdiag}}
\end{figure}

This relation leads to quantitative bounds on the sensing capacity of
equilibrium networks.  In general, the variance $\sigma_{RL}^2$, and
hence $\left( \delta c / c \right)^2_{RL}$, depends on the particular
network.  However, for any network, $\sigma_{RL}^2 \leq R_T^2/4$ since
$0 \leq RL \leq R_T$.  Thus, for equilibrium systems, the
  fundamental lower bound on the fractional error in the concentration
  estimate is:
\begin{equation}
\left( \frac{\delta c}{c} \right)_{x}^2 \ge \frac{4}{R_T^2}.
\label{eq:bound_eq_net}
\end{equation}
This proves that in equilibrium systems, which are not driven by fuel
turnover, the precision of sensing is fundamentally limited by the
number of receptors (Fig. \ref{fig:eqneqdiag}, upper box); a downstream signaling
network can never improve the accuracy of sensing.

Networks in which the receptors cooperatively bind the ligand can
  achieve the bound of Eq. \ref{eq:bound_eq_net} (\emph{SI Text}).  For networks without
  cooperative ligand binding, as in the simple example above, the
  sensing error is worse: $\sigma^2_{RL} \leq {\rm MIN}(R_T,R_T^2/4)$,
  so $\left( \frac{\delta c}{c} \right)_{x}^2 \ge {\rm
    MAX}(\frac{1}{R_T},\frac{4}{R_T^2})$ (\emph{SI Text}).  The
  sensing error for independent receptor binding is most easily
  understood for receptors with identical affinity for the ligand, as
  in our simple example (Eq. \ref{eq:receperr}), but holds generally:
  different affinities do not break this bound.

The different species in a network can also be viewed as nodes through
which information about the ligand flows. We can show that the data
processing inequality (\citealp{Cover2012}) also guarantees, for an equilibrium
system, that no read-out has more information about the ligand
 than the receptors at any given time: $I($x$; \mu_L)
\le I($RL$; \mu_L) \le \log_2 (R_{\rm T} + 1)$, where $I$ is the
mutual information between the instantaneous levels of the arguments
({\em SI Text}).  The history of receptor states does contain more
information about the ligand concentration than the instantaneous
receptor state, but our results show that an equilibrium signaling
network cannot exploit this: its output contains only as much
  information as the instantaneous receptor state; it does not encode
the history of receptor states in any informative way, whether by
time-integration or any other method.

Ultimately, equilibrium systems sense by harvesting the energy of
ligand binding. This energy is used to propagate the signal through
the downstream network; in the simple system studied here, for
example, the energy of ligand binding is used to expel the read-out
molecule from the receptor. However, detailed balance then dictates
that the receptor-read-out binding also influences receptor-ligand
binding, thus perturbing the signal. Indeed, the trade-offs faced by
equilibrium networks are all different manifestations of their
time-reversibility (\citealp{Feng2008}).  The only way for a time-reversible system to
``integrate'' the past is for it to integrate---perturb---the
future. Concomitantly, in a time reversible system, there is no sense
of ``upstream'' and ``downstream'', concepts which rely on a direction
of time.  Although we have referred to the molecule $x$ as a
``readout'' of the ligand concentration, the ligand is just as much a
readout of $x$.  While in equilibrium systems the
read-out encodes the receptor state, the read-out is not a stable
memory that is decoupled from changes in the receptor state. It merely passively lags.  In an
equilibrium system, the sensing error, like any static quantity, can only
depend on ratios of time scales,
which is another way of seeing that increasing the ``integration
time'' cannot improve sensing.

These results show that in an
equilibrium system each
receptor provides at most one independent measurement of the ligand,
regardless of how much information is encoded in the history of the
receptor state, how complicated the signaling machinery is downstream,
how many molecules are devoted to signaling downstream, or how long
the apparent integration time of the network is.    Energy
dissipation---fuel turnover---is required to break the trade-offs between noise
and sensitivity, between intrinsic and extrinsic noise, and,
ultimately, between the accuracy of sensing and space on the membrane.

\section{Non-equilibrium sensing at the molecular level}
Networks that can reduce error via time-integration
 must be non-equilibrium
systems.   To  understand the resources required to reduce the sensing error in these systems, we need to understand how they sense at the
molecular level.  Berg and Purcell pointed out that by integrating the
receptor signal over a time $T$, the cell can take as many as $T/(2 \tau_c)$ independent samples
of the receptor state (\citealp{berg1977}), where $\tau_c$ is the receptor correlation time.  We will show that the cost of sensing depends on how
many of these samples the cell actually takes. We therefore view the
downstream network, which consists of discrete components, as a system
that discretely samples the receptor state, rather than integrating it.  

\begin{figure}
\centering
\includegraphics[width=8.5cm]{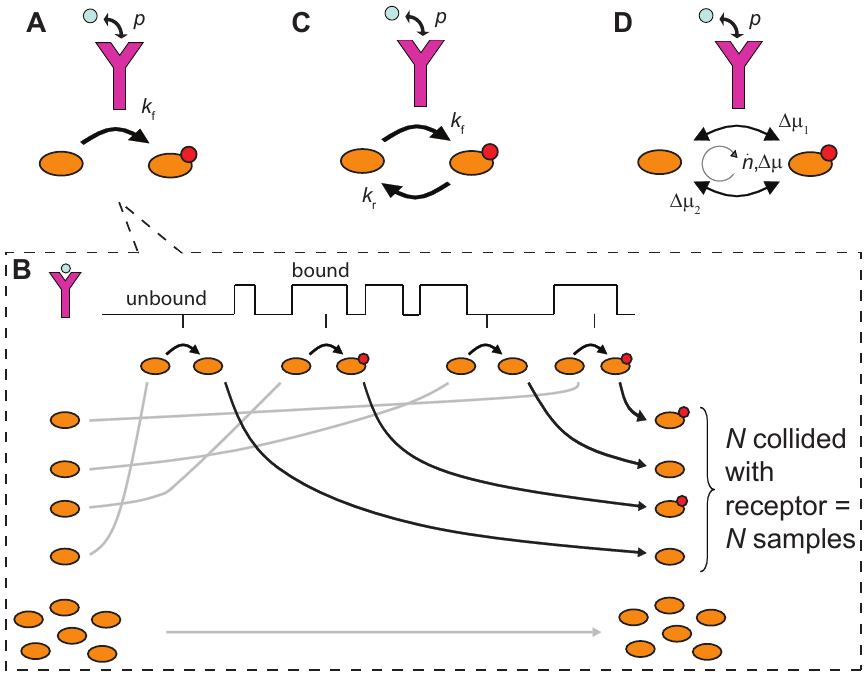}
\caption{ Sensing at the molecular level. (A) The
ligand-bound receptor drives the modification of a downstream readout
(i.e the push-pull network $RL + x \to RL + x^*$).  (B) The
biochemical network in (A) discretely samples the receptor state,
illustrated for one receptor.  The states of the receptor over time
are encoded in the states of the $N$ molecules that collided with it:
the readout is modified if the receptor is bound; otherwise it is
unmodified.  Molecules that collide with the unbound receptor are
indistinguishable from those that have never collided, leading to an
additional error.  (C) Active molecules can be degraded.  Some samples
are erased, and the remaining samples are, on average, further apart
(more independent.)  (D) All reactions are in principle reversible,
compromising the encoding of the receptor state into the readout. The
sensing error is determined by parameters that describe the energy flow
in the network, including the flux $\dot{n}$ and the free-energy drops
$\Delta \mu_1$ and $\Delta \mu_2$ across the activation and
deactivation reactions of the readout, respectively.\label{fig:sensing}}. 
\end{figure}

To gain intuition about the resources 
required to build and operate these networks,
we construct step by step a model of a receptor that drives a
push-pull network, which is a canonical non-equilibrium motif in prokaryotic and
eukaryotic cell signaling (\citealp{Alon:2007tz}). In these systems, the
receptor itself or the enzyme associated with it, such as CheA in
bacterial chemotaxis (\citealp{Alon:2007tz}), catalyzes the (chemical)
modification of a read-out protein. The general principle is that 
these networks take samples of the receptor by storing its
state in the stable modification states of the read-out molecules (Fig. \ref{fig:sensing}A, B).  Each read-out
molecule that interacted with the receptor provides a memory of the ligand-occupation state of that
receptor molecule; collectively, the read-out
molecules encode the history of the receptor states.  Quantitatively, if there are $N$ receptor-readout interactions,  then the cell has $N$ samples of the receptor state and the error, $\delta c/c$, is reduced by a factor of $\sqrt{N}$, as in Eq. \ref{eq:receperr}, or less if the samples are not independent.  By building up
the model step by step, we seek to understand how different features
of the network affect the number of samples, their independence, and
their accuracy.  One feature is that molecules can be deactivated (Fig. \ref{fig:sensing}C),
which we will show is equivalent to discarding or erasing  samples. Additionally,
reactions are microscopically reversible (Fig. \ref{fig:sensing}D), which means that read-out
modifications can occur independently of the receptor and
receptor-mediated modifications can occur in the wrong direction; both
effects reduce the reliability of a sample. Energy is needed to break
time-reversibility and to protect the coding.  We arrive at an
expression for the sensing error that combines these effects.  It
reveals trade-offs between cellular resources and performance: speed,
accuracy, energy, and the number of receptor and downstream molecules.

\subsection{Base model}

For intuition, we first consider a cell that responds
after a time $T$ to a change in a ligand's concentration at some time
$t=0$, based on the output $x^*$ of the simple reaction network $x +
RL \overset{k_{\rm f}}\to x^* + RL$ (Fig. \ref{fig:sensing}A).   We
assume that the cell starts with a large pool of inactive read-out
molecules $x$ and that activated molecules $x^*$ are never deactivated.
  For descriptive ease, we assume the
reaction is diffusion-limited, so that each collision between an
inactive molecule $x$ and a ligand-bound receptor leads to
activation of $x$.   The resulting sensing error can be derived via Eq. \ref{eq:error} from the Master Equation, which describes fluctuations in the network (see Materials and Methods).  

However, to understand the required resources, we calculate the error instead by viewing the molecular network as one that discretely samples the receptor state.  At the molecular level, readout molecules collide with the receptor over time and are modified depending on the ligand-occupation state of the receptor.  The total rate at which inactive
molecules collide with receptor molecules in any state is $r = k_{\rm f} x R_T
\approx k_{\rm f}
X_T R_T$  for a large readout pool, and the total number of such collisions after time $T$ is
$N$, with $\bar{N}=rT$ on average.   If a receptor molecule is bound to
ligand at the time of a collision, the read-out molecule is converted
to its active form, while if it is not the read-out remains unchanged.
In this way, the state of the receptor at the time of a collision is
encoded in the state of the read-out molecule that collided with it,
and the history of the receptor states is encoded in the states of the
read-out molecules at the time $T$
(Fig. \ref{fig:sensing}B). The read-out molecules that collided with
the receptor thus constitute samples of the receptor state. The
average number of samples after time $T$ is $\bar{N}= rT = k_{\rm f} X_T R_T T$---the product of the
total number of receptors $R_T$ and the number of samples per
receptor $k_{\rm f} X_T T$ during the integration time $T$.  

The sensing error can then be derived by viewing the system as one  that employs the sampling process described above, estimating the average receptor occupancy from samples of the receptor state taken at the times of readout-receptor collisions, {\em i.e.} as $\hat{p}=x^*/\bar{N}$ ({\em SI Text}).  This yields for the sensing error:
\begin{equation}
\label{eq:twoterm}
\left( \frac{\delta c}{c} \right)^2 = \frac{1}{p (1-p)}
\frac{1}{\bar{N}_I} + \frac{1}{(1-p)^2} \frac{1}{\bar{N}}.
\end{equation}
This expression has a clear interpretation in terms of sampling.  The first term is exactly the error expected from $\bar{N}$
stochastically taken samples of the receptor over the time
$T$. Specifically, it is the error of an estimate based on a single sample, $1/(p(1-p))$, divided by the average number of 
 samples that are independent, $\bar{N}_I$, where $\bar{N}_I$ is the total number of
samples $\bar{N}$ times the fraction $f_I$ that is independent:
\begin{equation}
\label{eq:ni}
\bar{N}_I = f_I \bar{N} = \frac{1}{1+\frac{2 \tau_c}{\Delta}} \bar{N},
\end{equation}
when $T \gg \tau_c$.  Clearly, $f_I$ depends on the receptor
  correlation time $\tau_c$ and on the time interval \linebreak $\Delta \equiv
  T/(\bar{N}/R_{\rm T})=1/(r/R_{\rm T})=1/(k_{\rm f} X_{\rm T})$
  between samples of the same receptor; samples farther apart are
more independent. This expression shows that the finite sampling rate
$r$ reduces the number of independent samples below the Berg-Purcell
factor $R_T T/2\tau_c$, the maximum number of independent samples that
can be taken during $T$. The latter is reached only when the sampling
rate is infinite (e.g. the number of downstream molecules $X_T\to
\infty$), so that $\bar{N}\to \infty$ and $\Delta \to 0$.

The second term in Eq. \ref{eq:twoterm} accounts for the fact that the
cell cannot distinguish between those molecules $x$ that have collided
with an unbound receptor (and hence provide information on the
receptor occupancy), and those that have not collided with the receptor at all (Fig. \ref{fig:sensing}B). If the cell could
distinguish between those molecules, it could
 estimate the average receptor
occupancy from $\hat{p}=x^*/N$ rather than $\hat{p}=x^*/\bar{N}$; then
the second term would be zero (\emph{SI Text}).  Indeed, the second term arises from the biochemical noise that makes the actual number of samples, $N$, different from its average, $\bar{N}$.  However, when $p$ is small and/or $\bar{N}$ is
large, the second term is small compared to the first and the sensing
error is given by the error of a single measurement, $1/(p(1-p))$, divided by the
average number of independent measurements, $\bar{N}_I$.

\subsection{Deactivation}
The error in Eq. \ref{eq:twoterm} decreases with the time $T$,
suggesting that the cell can sense perfectly if it waits long enough
before responding to a change in its environment.  However,
modification states of molecules decay, and their finite
lifetime, $\tau_{\ell}$, limits sensing, regardless of how long
the cell waits. To explore this at the molecular level, we consider
the network in the previous paragraph augmented with the deactivation
reaction $x^* \overset{k_{\rm r}} \to x$, $k_{\rm r} = 1/\tau_{\ell}$ (Fig. \ref{fig:sensing}C).  We consider the sensing error after long times
($T \gg \tau_{\ell}$), in steady state, again for a large pool of inactive
 read-out molecules.  For pedagogical clarity, we imagine the deactivation is
mediated by a phosphatase and that the reaction is diffusion-limited. 

We calculate the sensing error by solving the master equation or by
viewing the system as one that discretely samples the receptor state,
as before ({\em SI Text}). We find that also with deactivation the
sensing error is given by Eqs. \ref{eq:twoterm} and \ref{eq:ni}, yet
with fewer samples, $\bar{N}=r \tau_{\ell}< rT$, spaced effectively
farther apart, $\Delta = 2 \tau_{\ell}/(\bar{N}/R_T) = 2/(k_{\rm f}
X_T) > 1/(k_{\rm f} X_T) $. The molecular picture of sampling provides
a clear interpretation. As before, the readout molecules encode the
state of a receptor and serve as samples of the receptor state.  With
deactivation, however, only those readout molecules which have
collided with the receptor more recently than with the phosphatase
reflect the receptor state.  At any given time, the average number of
such readout molecules, and hence samples, is $\bar{N} = r
\tau_{\ell}$; the lifetime $\tau_{\ell}$ thus sets an effective
integration time. As without deactivation, the fraction $f_I$ of
samples that are independent is determined by the effective spacing
$\Delta$ between them, see Eq. \ref{eq:ni}. Though the time between
the creation of samples is still $1/(k_{\rm f} X_T)$, {\it i.e.} the
spacing without readout deactivation, some of the samples are erased
via collision with the phosphatase.  We therefore expect that the
spacing between remaining samples is larger.  Indeed, calculating the
effective spacing between samples taking this effect into account
yields $\Delta = 2/(k_{\rm f} X_T)$, which is twice that without decay
({\em SI Text}).  The fact that the remaining samples are more
independent explains a previously noted correspondence
(\citealp{Mehta2012,wingreen2009}) between the sensing error in a
  system with
deactivation, $\left( \frac{\delta c}{c} \right)^2 = \frac{1}{p
    (1-p)} \frac{1}{\tau_{\ell}/ \tau_c}$, and that in a system without deactivation, $
\left( \frac{\delta c}{c} \right)^2 = \frac{1}{p (1-p)} \frac{1}{T/(2
  \tau_c)}$, in the infinite sampling limit: they are equal for $T = 2
\tau_{\ell}$, and not for $T = \tau_{\ell}$ as would
be expected if their samples were just as independent.

\subsection{Finite pool of read-out molecules}
 The copy numbers of
signaling molecules are often small. To take this into account,  we compute the sensing
error from Eq. \ref{eq:error} for a finite number of read-out
molecules $X_{\rm T}$ using the linear-noise approximation to the Master Equation describing the biochemical fluctuations ({\em
  Materials and Methods}), and compare the
result with Eq. \ref{eq:twoterm}. This defines an effective number of
samples, $\bar{N} = r \tau_r$, where $\tau_r $ is the relaxation
time of the network.  For this network, $\tau_r = 1/(k_{\rm f} p R_T +
k_{\rm r})$.  In essence, cells count only
those samples created less than a relaxation time in the past; nothing
that happened earlier can influence the current state, including its
ability to sense.  The fraction of samples that is independent
is given by Eq. \ref{eq:ni} with $\Delta = 2
\tau_r/(\bar{N}/R_T)=2/(k_{\rm f} \bar{x})$, analogously to the previous section.

\subsection{Reversibility}
 All reactions are in principle
microscopically reversible.   Taking
this into account, we recognize that active molecules
that collide with the bound receptor sometimes become inactive, $x^* +
RL \to x + RL$, and that inactive molecules that collide with the
phosphatase are sometimes activated, $x \to x^*$ (Fig. \ref{fig:sensing}D).  These reverse
reactions compromise the encoding of the receptor state into the
read-out: an active $x^*$ molecule no
longer encodes the ligand-bound state of the receptor at a previous
time with 100\% fidelity, since it can also result from a collision
with the phosphatase; similarly, $x$, rather than $x^*$, may reflect
a collision with the ligand-bound receptor.

We compute the sensing error for the reversible network from
Eq. \ref{eq:error} using the linear-noise approximation to the master equation (see \emph{Materials and Methods}). As before, it
 can be written as
Eqs. \ref{eq:twoterm} and \ref{eq:ni}.  The effective number of independent samples $\bar{N}_I$
 is a complicated expression of the 8 fundamental variables
in the system: the 6 rate constants describing the forward and reverse
rates of the 3 reactions (including ligand-receptor binding), and the
total copy numbers $X_T$ and $R_T$.  However, the expression has a
particularly simple and illuminating form in terms of variables that describe, as
we will show, the resource limitations of the cell.  In addition to
variables already defined ($p$, $\tau_c$, $R_T$, and $\tau_r$), these include: the flux
$\dot{n}$ of $x^*$ across the cycle in which it is created
by the receptor and deactivated via the phosphatase; and the
average free-energy drops, $\Delta \mu_1$ and $\Delta \mu_2$, across
the receptor-catalyzed pathway and the phosphatase-catalyzed pathway,
respectively, in units of ${\rm k_ B T}$ (Fig. \ref{fig:sensing}D).  Each of these variables depends in a
complicated way on the fundamental parameters of the system, the rate
constants and the copy numbers. In particular, the free-energy
drops are related to the propensities 
$\nu_i$ and $\nu_{-i}$ of the reactions in the forward and backward
directions, respectively: $\Delta \mu_i = \log \frac{\nu_i}{\nu_{-i}}$
(\citealp{Seifert2012}).  However, the variables
can all be varied independently, except that $\Delta \mu_1 =
  \Delta \mu_2 = \dot{n} = 0$ in equilibrium.

In terms of these variables, the effective number of independent samples taken by
the push-pull network is:
\begin{equation}
\label{eq:neff}
\bar{N}_I =  \underbrace{\overbrace{\frac{\dot{n}
      \tau_r}{p}}^{\bar{N}} \overbrace{\frac{\left( e^{\Delta \mu_1} -
        1 \right) \left( e^{\Delta \mu_2} - 1 \right)}{ e^{\Delta \mu}
      - 1}}^{q}}_{\bar{N}_{\rm eff}} \underbrace{\frac {1}{(1+2\tau_c/\Delta)}}_{f_I},
\end{equation}
where $\Delta \mu=\Delta \mu_1+\Delta \mu_2$ is the total free-energy drop
across the cycle; $e^{\Delta \mu}$ is also known as the affinity of the cycle (\citealp{Schnakenberg1976}).  

Eq. \ref{eq:neff} is our principle result for non-equilibrium systems.
It takes into account readout deactivation, the finite number of
readout molecules, and the reversibility of reactions. The equation
has a clear interpretation.  The product $\dot{n} \tau_r$ is the
number of cycles of read-out molecules involving collisions with
ligand-bound receptor molecules during the system's relaxation
  time $\tau_r$. The quantity $\dot{n} \tau_r/p$ is the total number
of read-out cycles involving collisions with receptor molecules, be
they ligand bound or not; it is thus the total number of receptor
samples taken during $\tau_r$, $\bar{N}$.  The factor $q$, involving
$\Delta \mu_1,\Delta \mu_2$, reflects the quality of each sample.
When $\Delta \mu=\Delta \mu_1=\Delta \mu_2=0$, an active read-out
molecule is as likely to be created by the ligand-bound receptor as by
the phosphatase and there is no coding and no sensing; indeed, in
  this limit, $q=0$ and the effective number of samples
    $\bar{N}_{\rm eff}=0$. Note also that when the backward reactions are
faster than the forward reactions, corresponding to $\Delta \mu$
becoming negative, $x$ encodes the ligand-bound receptor instead of
$x^*$.  This symmetry is reflected by the symmetry of
Eq. \ref{eq:neff}: the number of samples is the same when the signs of
$\dot{n}$, $\Delta \mu_1$, and $\Delta \mu_2$ are all flipped.  The
effective number of accurate samples is $\bar{N}_{\rm eff} = \bar{N}
q$, less than the total number $\bar{N}$ taken.  The fraction of the
$\bar{N}_{\rm eff}$ samples that are independent is, as before,
$f_I=1/(1+2\tau_c/\Delta$) with $\Delta = 2
\tau_r/(\bar{N}_{\rm eff}/R_{T})$ reflecting the time interval between
effective samples.

\section{Trade-offs in non-equilibrium sensing}
Eq. \ref{eq:neff} reveals trade-offs  among the different resources for sensing, and between these resources and the accuracy of sensing.

\begin{figure}
\centering
\includegraphics[width=8.5cm]{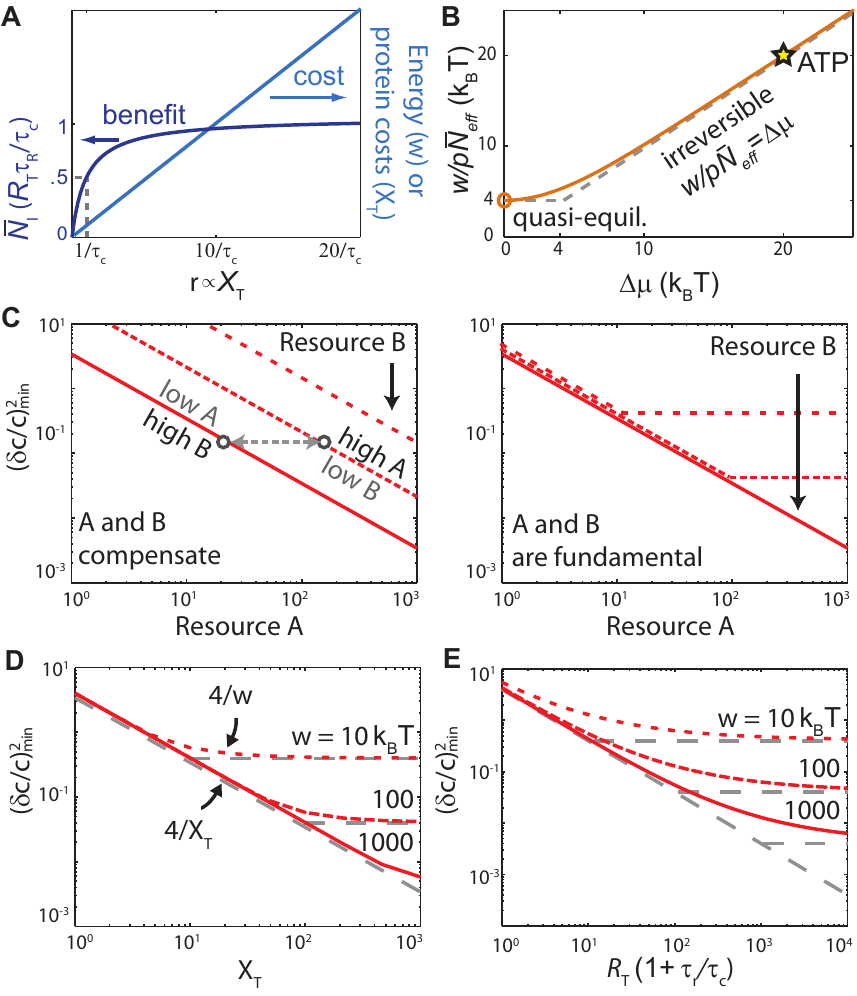}
\caption{Trade-offs in non-equilibrium sensing.  (A)
Sampling more than once per correlation time requires more resources,
while the benefit is marginal.  As the sampling rate $r$ is increased
by increasing the readout copy number $X_T$, the number of independent
measurements $\bar{N}_I$ saturates at the Berg-Purcell limit $R_T
\tau_r/\tau_c$, but the energy consumption and protein cost ($\propto
X_T$) continue to rise.  (B) The energy requirements for sensing.  In
the irreversible regime ($\Delta \mu > 4 {\rm k_BT}$), the work to
take one sample of a ligand-bound receptor, $w/(p\bar{N}_{\rm eff})$,
equals $\Delta \mu$, because each sample requires the turnover of one
fuel molecule, consuming $\Delta \mu$ of energy.  In the
quasi-equilibrium regime ($0<\Delta \mu < 4 {\rm k_BT}$), each
effective sample of the bound receptor requires $4 {\rm k_BT}$, which
defines the fundamental lower bound on the energy requirement for
taking a sample.  When $\Delta \mu = 0$, the network is in equilibrium
and both $w$ and $\bar{N}$ are $0$. ATP hydrolysis provides $20 {\rm
  k_BT}$, showing that phosphorylation of read-out molecules makes
  it possible to store the receptor state reliably. The results are
obtained from Eq. \ref{eq:neff} with $\Delta \mu_1 = \Delta \mu_2 =
(1/2) \Delta \mu$. (C) When two resources A and B compensate each
other, one resource can always be decreased without affecting the sensing
error, by increasing the other resource; concomitantly, increasing a
resource will always reduce the sensing error.  When both resources
are instead fundamental, the sensing error is bounded by the limiting
resource and cannot be reduced by increasing the other resource.  (D,
E) The three resources time/receptor copies, copies of downstream
molecules, and energy are all required for sensing, with no trade-offs
among them (see Fig. \ref{fig:eqneqdiag}B). The minimum sensing error
obtained by minimizing Eq. \ref{eq:twoterm} is plotted for different
combinations of (D) $X_T$ and $w$, and (E)
$R_T(1+\tau_r/\tau_c)$---the number of samples were $X_T$ and $w$
infinite---and $w$.  The curves track the bound for the limiting
resource indicated by the grey lines, showing that the resources do
not compensate each other. The plot for the minimum sensing error as a
function of $R_T(1+\tau_r/\tau_c)$ and $X_T$ is identical to that of
(E) with $w$ replaced by $X_T$.
  \label{fig:efficiency}}
\end{figure}

\subsection{Time/receptor copy numbers}
There is no fundamental
relationship between receptor copy number and sensing, as in
equilibrium systems.  Essentially, the error is determined by the
total number of samples, and it does not matter, as long as the
samples are independent, whether these samples are from the same
receptor over time or from many receptors at the same time. An
independent sample of the same receptor can be taken roughly every $2 \tau_c$
(Eq. \ref{eq:ni}). Naturally, samples can be taken more frequently. In
fact, cells can time-integrate: if $X_{\rm T}\to \infty$, the
receptors are sampled infinitely fast and  $\Delta \to 0$ and
$\bar{N}_{\rm eff}\to \infty$ (Eq. \ref{eq:ni}); then the number of independent
samples $\bar{N}_I$ taken over $\tau_r$ reaches its maximum, the Berg-Purcell
factor $R_T \tau_r/\tau_c$, and the error can achieve its minimum, $4/(R_T \tau_r/\tau_c)$. However, the benefit of sampling faster by increasing  $X_{\rm T}$ in
reducing the sensing error is rapidly diminishing, while the total protein and energetic
costs increase (see Fig. \ref{fig:efficiency}A).

\subsection{Downstream read-out molecules}
While the effect of copy number on intrinsic noise has been
studied extensively, how copy numbers of read-out signaling molecules
affect the fundamental sensing limit has not been elucidated. In
Eq. \ref{eq:neff} the factor containing the chemical potentials is
always less than 1; also, $\dot{n} \tau_r < X_{T}$ because the system
has relaxed when all read-out molecules have completed an
activation-deactivation cycle. Hence, Eq. \ref{eq:neff} shows that the
number of samples of a ligand-bound receptor, $p\bar{N}$, is always
less than the number of downstream molecules, $X_{T}$. Each read-out
molecule provides at most one sample, because at any given time it
exists in only one modification state, regardless of how many times it
has collided with the receptor or how long the integration time
$\tau_r$ is.  There is no mechanistic sense in which a single molecule
"integrates" the receptor state.  As a consequence, no matter how the
network is designed, how much time or energy it uses, or how many
receptors it has, cells are fundamentally limited by the pool of
read-out molecules: the sensing error $(\delta c/c)^2 \geq 4/X_{T}$,
obtained by analytically minimizing Eq. \ref{eq:twoterm} at fixed $X_T$.  

\subsection{Energy}
The free-energy drop across a cycle, $\Delta \mu$,
must be provided by a fuel molecule such as ATP.  This free energy
represents the maximum work the fuel molecule could do if used by an
ideal engine.
Eq. \ref{eq:neff} defines three regimes of sensing with respect to the
energy consumption of the network.  

When $\Delta \mu=0$ the system is in equilibrium and the sensing error
diverges (\citealp{Mehta2012}), as discussed above; indeed, this
system employs a fundamentally different signaling strategy than
equilibrium systems use to sense.  Two other regimes are defined by
the work that the fuel molecules need to do in order to take a
  sample of the receptor.  The power, the rate at which the fuel
molecules do work, is $\dot{w}=\dot{n}\Delta \mu$ and the total work
performed during the relaxation time $\tau_r$ is $w\equiv \dot{w}
\tau_r$. This work is spent on taking samples of receptor
  molecules that are bound to ligand, because only they can modify
  downstream read-out molecules. The total number of effective samples
  of ligand-bound receptors obtained during $\tau_r$, is
  $p\bar{N}_{\rm eff}$.  Hence, the work needed to take one  effective
sample of a ligand-bound receptor is $w/(p\bar{N}_{\rm eff})$, with
$\bar{N}_{\rm eff}$ given by
Eq. \ref{eq:neff}. Fig. \ref{fig:efficiency}B shows this quantity as a
function of $\Delta \mu$. Two regimes can be observed.

When $\Delta \mu \gtrsim 4 {\rm k_B T}$, the work to take one
effective sample of a ligand-bound receptor is simply $w
/(p\bar{N}_{\rm eff})=\Delta \mu$, independent of kinetic time scales.
In this regime, the read-out reactions are essentially irreversible
and the sample quality factor $q$ in Eq. \ref{eq:neff} reaches unity,
meaning that each read-out molecule reliably encodes the receptor
state at an earlier time.  The effective number of samples therefore
equals the total number taken, and is given by that of the
irreversible case already studied, $\bar{N}_{\rm eff} =\bar{N} =
\dot{n}\tau_r/p=r \tau_r = k_{\rm f} \bar{x} R_{T} \tau_r$.  The work
per sample of a ligand-bound receptor, $w/(p\bar{N}_{\rm
  eff})$, equals $\Delta \mu$, because each sample requires the
turnover of one fuel molecule, using $\Delta \mu$ of energy. The total
number of samples $\bar{N}_{\rm eff}$ is thus limited by the work as
$\bar{N}_{\rm eff} = \dot{n}\tau_r/p = w / (p \Delta \mu)$. In this
regime, energy limits sensing not because it limits the reliability
$q$ of each sample, but because it limits the total number of samples
$\bar{N}_{\rm eff}$ that could be taken during $\tau_r$ by limiting
the receptor sampling frequency, the flux $\dot{n}$: increasing
$\dot{n}$ necessarily requires more work. 
Inserting this expression into Eq. \ref{eq:twoterm} and optimizing
puts a lower bound on the sensing error:
\begin{equation}
\left( \frac{\delta c}{c} \right)^2 \geq
\frac{1}{\dot{n}\tau_r}=\frac{\Delta \mu}{w}.
\label{eq:fundlim_irr}
\end{equation}
  Intriguingly,
Eq. \ref{eq:fundlim_irr} suggests that for a fixed amount of energy,
$w$, spent during the relaxation time $\tau_r$, the sensing error can
be reduced to zero by reducing $\Delta \mu$ to zero. However, the
lower bound in Eq. \ref{eq:fundlim_irr} is only achievable (and
Eq. \ref{eq:fundlim_irr} thus
only applies), when $\Delta \mu \gtrsim 4 {\rm k_B T}$.

When $\Delta \mu \lesssim 4 {\rm k_B T}$, the system
  transitions to a quasi-equilibrium regime in which each fuel
  molecule provides a small but nonzero amount of energy.  In this
  regime, the system can still consume significant amounts of energy
  when the fuel molecules are consumed at a rapid rate $\dot{n}$ by
  many distinct read-out molecules.  In the limit that $\dot{n} \to
  \infty$ and $\Delta \mu \to 0$ at fixed $\dot{w}=\dot{n}\Delta \mu$,
  the effective number of samples given by Eq. \ref{eq:neff} reduces
  to
\begin{equation}
\label{eq:fundlim}
\bar{N}_{\rm eff} \to \frac{w}{4p}.
\end{equation}
In the quasi-equilibrium regime,  each readout-receptor interaction
corresponds to an increasingly noisy measurement of the receptor
state ($q \to 0$), but many noisy measurements ($\bar{N} = \dot{n} \tau_r/p \to \infty$)
contain the same information as 1 perfect measurement -- provided that
collectively at least $4{\rm k_B T}$ was spent on them.  Indeed, as
Fig. \ref{fig:efficiency}B shows, $4{\rm k_B T}$ is the fundamental
lower bound on the work needed to take one accurate sample of a ligand-bound
receptor. It puts another lower bound on the sensing error: inserting
Eq. \ref{eq:fundlim} into Eq. \ref{eq:twoterm} and optimizing shows
that:
\begin{equation}
\left( \frac{\delta c}{c} \right)^2 \geq \frac{4}{w}.
\label{eq:ener_req} 
\end{equation}
This power-law bound relates energy to
information. The bound can be reached when time and $X_T$ are not
  limiting, and $\Delta \mu\lesssim 4 {\rm k_B T}$. When $\Delta \mu
  \gtrsim 4{\rm k_B T}$,  the lower bound is higher and given by
  Eq. \ref{eq:fundlim_irr}.

Eqs. \ref{eq:fundlim_irr} and \ref{eq:ener_req} show that the sensing
precision increases with the work done in the past relaxation time,
$w=\dot{w}\tau_r$, setting up a trade-off among speed, power, and
accuracy, as found in adaptation (\citealp{Tu2012}).  The trade-off
emerges naturally from a molecular picture of sensing.  When the
response needs to be rapid, $\tau_r$ needs to be small and the power
demand is high: the samples, which require energy, must be taken close
together in time.  However, when the cell can wait a long time
$\tau_r$ before responding, the power $\dot{w}$ required to make $w$
large can be infinitesimal: the samples can be created far apart in
time. There is no minimum power requirement for sensing.

\subsection{No trade-offs among time/receptors, readout molecules, and energy}
The above analysis shows that each of the fundamental resource
categories --- time/receptor copy number, downstream read-out
molecules, and power/time (fuel) --- has its own trade-off with sensing
accuracy (Figs.~\ref{fig:efficiency}C,D,E).  To a good approximation,
the worst bound is active: $\bar{N}_I \leq {\rm MIN} \left( R_T\tau_r/
  \tau_c, X_T/p, \dot{w}\tau_r/4 p \right)$, corresponding to $(\delta
c/c)^2 \geq {\rm
  MAX}\left(4/(R_T\tau_r/\tau_c),4/X_T,4/(\dot{w}\tau_r)\right)$.
Indeed, one of the most important conclusions of our analysis is that
increasing a single resource (e.g. $w$) cannot reduce the sensing
error indefinitely.  The sensing accuracy will eventually plateau,
namely when it becomes fundamentally limited by another resource
(e.g. $X_T$).  Clearly, there is no trade-off among these classes of
resources: no amount of one resource can overcome a limiting amount of
another, as illustrated in Figs. \ref{fig:efficiency}D,E.  The reason is clear:
taking a sample requires time and receptors, read-out molecules, and
fuel. Adding receptors and read-out molecules does not improve
  sensing if not enough energy is available to take the samples 
(Fig. \ref{fig:efficiency}D). Similarly, waiting more time to take
another sample is not beneficial if the cell has no more read-out
molecules left to write the sample to, or cannot expend energy fast
enough to accomplish the writing (Fig. \ref{fig:efficiency}E).

The picture that emerges from our analysis is summarized in the lower box of
  Fig. \ref{fig:eqneqdiag}. The resource classes time/receptors,
  downstream readout molecules, and energy, act like weak links in a
  chain that cannot compensate each other in achieving a required
  sensing precision. Within these classes
  trade-offs are possible: time can be traded against the number of
  receptors to reach a required number of measurements, while power
  can be traded against speed to meet the energy requirement for a
  desired sensing accuracy.

\section{Trade-offs between different modes of sensing}

\begin{figure}[b]
\centering
\includegraphics[width=8.5cm]{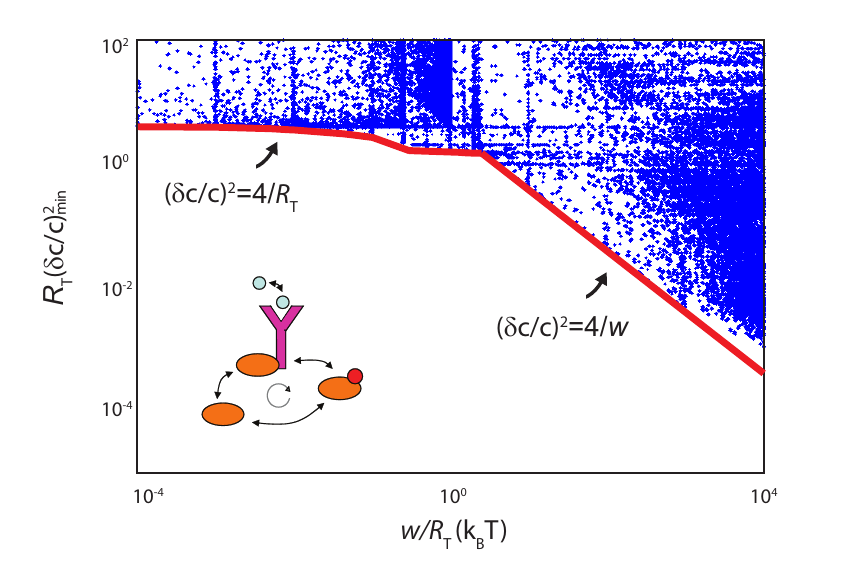}
\caption{ The different resource requirements for
  equilibrium and non-equilibrium sensing lead to a trade-off between
  these two modes. The trade-off is illustrated for a network that
  combines both modes:  $RL + x \rightleftharpoons RLx$, $RLx
  \rightleftharpoons RL + x^*$, $x^* \rightleftharpoons x$.  The blue
  dots show the sensing error for different parameter values and the red
  guideline shows the minimum sensing error ({\it SI
    text}). 
When the energy per receptor $w/R_T$ is less than a few ${\rm k_BT}$, the optimized
system employs the equilibrium strategy of sequestration, achieving the bound
  $4/R_T$.  If the energy input is higher, it uses the non-equilibrium strategy of catalysis to transmit
the signal, achieving the bound $4/w$.  There is an intermediate regime around $1$ ${\rm k_BT}$
  per receptor in which the network modestly outperforms both full catalysis
  and full binding by partially utilizing the receptor-read-out state.\label{fig:eqneqmin}}
\end{figure}

To increase the number of measurements, equilibrium networks must
increase the number of receptors.  Non-equilibrium networks may
instead use more downstream readouts and energy to take more
measurements with the same receptors over time.  Which sensing
strategy is better?  The strategy adopted by the cell will depend on
the relative fitness costs of the different resources. If the
resources are of similar costs, then, quantitatively, our
analysis predicts that an equilibrium strategy will be adopted if its
minimum error, $4/R_T$ for non-cooperative receptors, is less than
that of the non-equilibrium strategy, ${\rm
  MAX}\left(4/(R_T\tau_r/\tau_c),4/X_T,4/(\dot{w}\tau_r)\right)$ (Fig. \ref{fig:eqneqdiag}).  For example,
when the accuracy of the non-equilibrium strategy is limited by
energy, meaning that $(\delta c /c)^2\geq 4/(\dot{w}\tau_r) $, the predicted
transition between the two strategies occurs when the work per
receptor $\dot{w}\tau_r/R_T
\approx 1 {\rm k_B T}$. To address this, 
we have considered a network that combines both strategies.  The
read-out binds the ligand-bound receptor, which can then boot off the
read out in a modified or unmodified state ({\em SI Text}): $RL + x
\rightleftharpoons RLx$, $RLx \rightleftharpoons RL + x^*$, $x^*
\rightleftharpoons x$. This system combines both modes of sensing,
because the chemical modification of the readout enables
non-equilibrium sensing, while sequestration of the unmodified readout
by the receptor upon ligand binding enables equilibrium sensing.
Optimizing this system over all parameters confirms that when the
energy per receptor is less than a few ${\rm k_BT}$, the optimized
system employs the equilibrium strategy of sequestration, while if it is
higher it uses the non-equilibrium strategy of catalysis to transmit
the signal (Fig. \ref{fig:eqneqmin}).  In addition, the networks that
optimize sensing in these two regimes are the networks that we have
studied; a network that combines the two sensing modes does not
perform better than the two individually.

\section{Discussion}

Fundamentally there are only two distinct mechanisms to transmit the
information from the receptor to the downstream readout (Fig. \ref{fig:eqneqdiag}).  These two mechanisms, described
as equilibrium and non-equilibrium sensing, have different resource
requirements (Table \ref{tab:comparison}). Cells face a trade-off with
respect to their resources in choosing between these two distinct sensing
strategies. 

In the equilibrium strategy the signal is transmitted from the
receptor to the read-out via sequestration reactions, in which binding
of an upstream component causes unbinding of a downstream component,
or via adaptor proteins, which bind the up- and downstream component
simultaneously. Both motifs are ubiquitous in cellular
signaling. G-protein-coupled receptor (GPCR) signaling employs
protein sequestration (\citealp{Neves:2002uq}), while Ras signaling
uses adaptor proteins like Grb2 (\citealp{Maignan:1995fk}).

\begin{table*}
\caption{Two strategies for sensing the environment}
\centering
\label{tab:comparison}
\begin{tabular}{  p{1.5in}  p{2in}  p{2.5in}   }
	\hline
	& \multicolumn{1}{c}{Equilibrium sensing} & \multicolumn{1}{c}{Nonequilibrium sensing}  \\[1ex]
	\hline 
  \emph{Example implementation} &  \multicolumn{1}{c}{\includegraphics{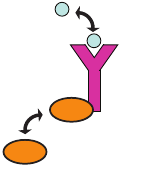}}  & \multicolumn{1}{c}{\includegraphics{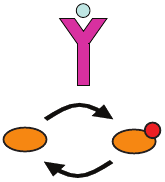}}  \\[.5ex]
  \emph{Biochemical mechanism} & sequestration/adaptor proteins & catalysis \\[.5ex]
  \emph{Sampling strategy} & measurements with more receptors & measurements of same receptors over time \\[.5ex]
 \emph{Fundamental resources} & receptors & receptors/time, fuel, readout molecules \\ [.5ex]
 \emph{Energy source} & harvests energy of ligand binding & energy from fuel \\[.5ex]
\emph{Intrinsic and extrinsic noise linked?} & yes & no \\[.5ex]
 \emph{Minimum error} & $4/R_T$ (without cooperativity) & ${\rm  MAX}\left(4/(R_T\tau_r/\tau_c),4/X_T,4/w\right)$ \\
 & $4/R_T^2$ (with cooperativity) &  \\[.5ex]
 \emph{Example} & one-component signaling & two-component signaling \\[.5ex]
\hline
\end{tabular}
\end{table*}

Equilibrium systems do not require fuel turnover. They respond to
changes in the environment by harvesting the energy of ligand binding,
thereby capitalizing on the work that is performed by the environment
to change the ligand concentration.  While the response speed is
determined by the rate constants, the accuracy of sensing is only
limited by their ratio; there is no trade-off between speed and
accuracy. The sensing precision is, however, limited by the number of
receptors. This is because the energy of receptor-ligand binding is
used to expel or bind the messenger protein, thus coupling
receptor-ligand binding to receptor-readout binding.  This inevitably
leads to correlations between the extrinsic noise in the receptor and
the intrinsic noise of the processing network
(\citealp{TanaseNicola2006}).  These correlations lead to a
fundamental trade-off between these sources of noise in equilibrium
systems.

In nonequilibrium sensing, fuel turnover allows the receptor to
transmit information as a catalyst. This makes it possible to
  remove the correlations and the concomitant trade-off between
  extrinsic and intrinsic noise, and
  reach a sensing precision that is not limited by the number of
  receptors.  Arguably the most important signaling motif that relies
  on fuel turnover is the Goldbeter-Koshland push-pull network studied
  here. This motif is used in most, if not all, signal transduction
  pathways.

  Our analysis reveals why it may be beneficial to use this energy
  consuming motif: it makes it possible to store
  the history of the receptor state in stable chemical modification
  states of downstream molecules. In equilibrium sensing the
    stability of the downstream signaling proteins relies on physical
    interactions with the receptor molecules, which means that the state of the readout
    molecules reflects the instantaneous state of the
    receptor. In contrast, in non-equilibrium sensing the energy to
    change the state of the signaling proteins is not provided by the
    physical interactions with the receptor, but by the chemical
    fuel. The receptor catalyzes the modification of the read-out. 
    After modification, however, the receptor and read-out become decoupled and each read-out molecule
    provides a stable memory of the receptor
    state when it was modified.   It is this feature
    that allows these non-equilibriums systems to take samples of the
    receptor state over time and perform a discrete time
    integration. This increases the number of measurements
    per receptor, making it possible to beat the equilibrium sensing
    limit set by the number of receptors. 

    Taking samples fundamentally requires time so that the samples are
    independent; downstream molecules to store the samples; and energy
    to store them reliably, to protect the coding.  We find that at
    least $4{\rm k_BT}$ is needed for reliable encoding, quantifying a
    relationship between energy and information.  One of the most
    widely used coding strategies is phosphorylation, which requires
    ATP. {\em In vivo}, ATP hydrolysis provides about $20 {\rm
      k_BT}$. This is sufficient to take one receptor sample
    essentially irreversibly (Fig. \ref{fig:efficiency}B), which means
    that the quality factor $q$ reaches unity. Readout phosphorylation
    thus makes it possible to store the receptor state reliably.

    Non-equilibrium networks can exhibit more complicated features
    than those of the simple push-pull motif, as in the MAPK
    cascade. The molecular picture for time-integration suggests that
    our results for the push-pull network hold generally, even in
    these more complicated systems.  Indeed, we find the same or more
    severe resource limitations in cascades and networks with simple
    negative or positive feedback ({\em SI Text}).  Although cascades
    can increase the response time (\citealp{Govern2012}), which
    increases information transfer, they do not make sensing more
    efficient in terms of energy or readout molecules.

    One- and two-component signaling networks provide a case study for
    the trade-off between equilibrium and non-equilibrium sensing.
    One-component systems consist of adaptor proteins which can bind
    an upstream ligand and a downstream effector, while two-component
    systems are similar to the push-pull network studied here,
    consisting of a kinase (receptor) and its substrate.
    Interestingly, some adaptor proteins, like RocR, contain the same
    ligand-binding domain as the kinase and the same effector-binding
    domain as the substrate of a two-component system, {\it i.e.}
    NtrB-NtrC (\citealp{Ulrich:2005ys}). It has been suggested that
    one-component systems have evolved into two-component systems to
    facilitate transfer of signals from the membrane to the nucleus
    (\citealp{Ulrich:2005ys}).  However, equilibrium networks can also
    transmit signals across space (Table \ref{tab:comparison}).  Our
    results thus suggest that these one-component systems are really
    alternative, equilibrium solutions to the problem of signal
    transduction, selected because of different resource selection
    pressures. 

    Which resource sets the fundamental limit for non-equilibrium
    sensing?  Although it has often been assumed that time/receptors
    are limiting (\citealp{berg1977,bialek2005, levinepre2007,
      levineprl2008,wingreen2009,levineprl2010,mora2010,Govern2012}),
    our results, in contrast, show how the accuracy of sensing can
    instead be limited by energy or downstream copy numbers.
    Interestingly, experiments suggest that some key networks are not
    time/receptor limited.  Cheong {\it et al.} have measured the
    information transmission of several important networks, and have
    shown that all transmit about 1 bit of information, or less
    (\citealp{Cheong:2011fk}).  This amount is far less than the
    networks would transmit if they were time/receptor limited (see
    {\em SI Text}). This suggests that another resource, such as copy
    numbers of signaling components or energy, limits sensing.  In
    such scenarios, characterizing the response time of the network is
    less important for understanding sensing than characterizing
    protein expression levels and their energy usage.

    It seems natural to expect that the resources which are limiting
    sensing are those that affect cell growth or fitness, while the
    resources that are in excess and thus wasted are those that do not
    significantly affect cell growth or fitness. This prediction could
    be tested experimentally, for example by studying the growth and
    chemotactic performance of bacterial populations with different
    expression levels of functionally and non-functionally signaling
    proteins (\citealp{Dekel2005}).  To the extent that all resources
    affect growth, evolutionary pressure should tend to drive systems
    so that no resource is wasted, which occurs when all are equally
    limiting.  Resource-optimal systems sample the receptor about once
    per correlation time and use just enough fuel and downstream
    molecules to do so. Quantitatively,
    all resources are equally limiting when
\begin{equation}
 R_{T}\tau_r/ \tau_c
    \approx X_{T} \approx w.
\label{eq:opt_sys}
\end{equation}
In an optimal sensing system, the number of independent
  concentration measurements $R_T \tau_r / \tau_c$ equals the number
  of readout molecules $X_T$ that store these measurements
  and equals the work (in units of ${\rm k_BT}$) to create the
  samples.

In two-component signaling systems, including that of bacterial
chemotaxis, the downstream component is typically in excess of the
  receptor (\citealp{Batchelor:2003fk,Falke:1997zr,Li:2004ly}).
For the  {\em E. coli} chemotaxis system, $X_T / R_T \approx 3-4$
(\citealp{Falke:1997zr,Li:2004ly}). Eq. \ref{eq:opt_sys} thus predicts that $\tau_r /
\tau_c \approx 3-4$. This prediction can be tested, assuming that the correlation time
$\tau_c$ of the receptor-CheA complex is that of receptor-ligand
binding.  In {\em E. coli}, the lifetime of the active
(phosphorylated) readout, CheYp, is $\tau_l \approx 100{\rm ms}$
(\citealp{Sourjik:2002fk}), which means that $\tau_r \approx \tau_l / 3 \approx 30
{\rm ms}$, since about a third of the total amount of CheY is
phosphorylated in steady-state. Eq. \ref{eq:opt_sys} thus predicts that $\tau_c
\approx 10 {\rm ms}$. To test this prediction, we estimate $\tau_c$
from the receptor-ligand dissociation rate $k_{\rm off}$ as $\tau_c
\simeq 1/(2 k_{\rm off})$, ($p\approx 0.5$). The dissociation constant
of Tar-aspartate (receptor-ligand) binding $K_D\approx 0.1-1\mu{\rm M}$ (\citealp{Vaknin:2007ys}) and
with an association rate $k_{\rm on}\approx 10^9 {\rm
  M}^{-1} {\rm s}^{-1}$ (\citealp{Danielson:1994vn}), this yields 
$k_{\rm off}\approx 100-1000 {\rm s}^{-1}$ and an estimated correlation
time $\tau_c \approx 1-10 {\rm ms}$, in line
with the prediction of Eq. \ref{eq:opt_sys}.

Eq. \ref{eq:opt_sys} also predicts that the fundamental resources should vary
    proportionally to each other across different networks.  For
    example, the relation predicts that the lifetime $\tau_r$ of the modified
    state of a readout molecule should increase, \emph{ceteris
      parabus}, with its expression level $X_T$.  Two-component systems can
    provide a large-data set for testing these predictions once
    kinetic data and protein expression levels for many of them become
    available (\citealp{Gao:2013vn}).

Our results are also important for synthetic biology, which uses
two-component signaling networks as a building block (\citealp{Ninfa:2010zr}).
The design principles  instruct how
such networks should be constructed at the molecular level to minimize
resource consumption.  In turn, synthetic networks may provide a
platform for testing key predictions.

A major question in cell signaling is to what extent the design of
signaling pathways is shaped by the same limits that apply to other
sorts of machines, and to what extent they face unique
limitations because they are constructed out of molecular networks.
The process of sampling a time series, like the receptor state over
time, defines a specific, familiar computation that could be conducted
by any machine; it is instantiated in the biochemical system by the
readout-receptor pair.  We find that the free-energy drops across the
``measurement'' and ``erasure'' steps, $\Delta \mu_1$ and $\Delta
\mu_2$, should be identical to minimize the energetic cost, even
though the fuel molecule need only be involved in one of the
reactions, preparing a non-equilibrium state that relaxes via the
other.  This allocation of energy differs from that typically
considered in the computational literature, in which only the erasure
step requires energy (\citealp{Landauer1961}).  In the cellular system both steps are
computational erasures: though only the ``erasure'' step erases memory
of the receptor state, both steps erase the state of the molecule
involved in the collision.  Interestingly, when $p=0.5$, the average work to
measure the state of the receptor is $2{\rm k_BT}$, which is perhaps
surprisingly close to the Landauer bound, ${\rm k_BT} \ln(2)$
(\citealp{Landauer1961}).

\section{Materials and methods}

\subsection*{Calculating the sensing error for a biochemical network}

From Eq. 1 in the main text, the sensing error for a biochemical
network depends on the gain and the variance of the readout molecule.
For all networks studied in this paper, we have calculated the gains
using a mean-field approximation for the steady state level of the
readout, which is exact for linear networks (the base model of the
main text and the base model plus deactivation).  Except where
otherwise noted, we have calculated all variances using a linear-noise
approximation (\citealp{Gardiner1985}), which is, again, exact for the
linear networks.  For nonlinear networks, the quality of the
approximation improves with system size; it can already be quite good
for systems with only 10 copies of each molecule
(\citealp{wiggins2007}).  For the base model and the base model with
deactivation, we have used tools of discrete stochastic
processes to independently calculate the error by viewing the signaling network as a
system that samples the receptor state (see \emph{SI Text})

The linear-noise approximation gives the covariance matrix $\Sigma$ for stationary fluctuations in species' levels as the solution to the Lyapunov equation:
\begin{equation}
A \Sigma + \Sigma A^T + B = 0 
\end{equation}
where $A = S^T \nabla \nu$  and $B = S^T \text{Diag} \left( \nu \right) S$ in terms of the stoichiometric matrix $S$ and the reaction propensity vector $\nu$.  The stoichiometric matrix describes how many molecules of each species are consumed or produced in each reaction, and the propensity vector describes the propensity (rate) of each reaction.  For a network out of steady state (the base model), a non-stationary version must be used (\citealp{Gardiner1985}).

\subsection*{Langevin approximation to the dynamics of a biochemical network}

The Langevin approximation to the dynamics of a biochemical network draws on the same framework as the linear-noise approximation (\citealp{Gardiner1985}).  It expresses the fluctuations in species copy numbers $\boldsymbol{N}$ as:
\begin{equation}
\frac{d\boldsymbol{N}}{dt} = A \boldsymbol{N} + \eta(t)
\end{equation}
where $\boldsymbol{N}$ is a vector containing the copy numbers of all species and $\eta(t)$ are Gaussian noises, uncorrelated in time, with covariance $B$.  $A$ and $B$ are the matrices defined in the section ``Calculating the sensing error for a biochemical network.''  The equation can be solved (e.g. by integrating factors; \citealp{Gardiner1985}), yielding the result in the main text, Eq. 3, for the biochemical network considered there.

\section{Acknowledgements}

 We thank Tom Shimizu, Andrew Mugler and Thomas Ouldridge for a critical reading
  of the manuscript. This work is part of the research programme of the
  Foundation for Fundamental Research on Matter (FOM), which is part
  of the Netherlands Organisation for Scientific Research
  (NWO).

\section{Supplementary Material} 

\section*{Minimum sensing error of the simple equilibrium binding system}

In the main text, we considered a simple equilibrium system in which the read-out binds the unbound receptor: $R + L \rightleftharpoons RL$, $R + x \rightleftharpoons Rx$.  Here, we show that the sensing error for this network is limited by the number of receptors on the surface of the cell, as stated in the main text.  Calculating the variance as described in the main text (or directly via the linear-noise approximation) yields for the sensing error:
\begin{equation}
\left( \frac{\delta c}{c} \right)^2 =  \frac{ (R + RL) (Rx(x) + R(Rx + x) + RL(Rx + x))}{(RL^2) (Rx) (x)} \nonumber
\end{equation}
where we have written the result (following the Lagrange multiplier approach from, for example, \citealp{Samuelson:1947kx}) in a form that makes it easy to show that the error is bounded by the number of receptors; a direct expression in terms of the rate constants is quite complicated.  Indeed, minimizing the expression over $R$, $Rx$, $RL$, and $X_T$ such that all are positive and $R+RL+Rx = R_T$ and $x+Rx = X_T$ shows the result in the main text, that the error is always greater than $\frac{1}{p(1-p)} \frac{1}{R_T}$ (Eq. 2 of the main text).

\section*{Cooperativity achieves the fundamental equilibrium bound and is necessary to achieve it}

First we show that cooperative binding of the ligand to the receptors can achieve the fundamental equilibrium bound.  One way in which receptors can cooperatively bind ligand is when the receptors are in clusters.  Consider $C_{\rm T}$ clusters, each containing $n$ receptors that cooperatively bind $n$ ligand molecules, $C + nL \rightleftharpoons CL^n$.  The number of bound clusters, $CL^n$, is binomially distributed, giving variance $C_T p (1-p)$, where $p$ is the probability a cluster is bound.  The fluctuation-dissipation theorem gives the gain as $n C_T p(1-p)$, since each cluster binds $n$ ligand molecules.  The sensing error is then (Eq. 1 in the main text): $\left( \frac{\delta c}{c} \right)^2 = \frac{1}{p (1-p)} \frac{1}{n^2 C_{\rm T}}$.   When all the receptors are in a single cluster ($n=R_T,C_T=1$), this can be as low as $4/R_T^2$, achieving the fundamental bound.

Equilibrium systems without positive cooperativity at the level of the receptors cannot achieve this bound, at least under the linear-noise approximation.  We prove this in the general case that multiple different receptor species, $R_i$, can bind the ligand, possibly with different affinities -- but not cooperatively.  The fluctuation-dissipation theorem guarantees that the best readout is the total number of bound receptors, $RL =  \sum R_iL$, since that is the variable conjugate to the chemical potential of the ligand.  In general, the variance $\sigma_{RL}^2$ is just the sum of the  variances of the species, plus corrections for the correlations between the species:
\begin{equation}
\sigma_{RL}^2 = \sum_i \sigma_{R_iL}^2 + \sum_i \sum_j \sigma^2_{R_iL,R_jL} \le \sum_i \sigma_{R_iL}^2
\end{equation}
where the inequality follows from the lack of (positive) cooperativity.  (Negative cooperativity can emerge naturally in equilibrium networks due to competition of downstream molecules for binding to the receptors.)  For an equilibrium system, the variance of a species is always less than the mean level of that species, at least under the linear-noise approximation, so:
\begin{equation}
\sigma_{RL}^2 \le \sum_i \overline{R_iL} \le R_T.
\end{equation}
Thus:
\begin{equation}
\label{eq:linearlim}
\left( \frac{\delta c} {c} \right)^2 = \frac{1}{\sigma_{RL}^2} \ge \frac{1}{R_T}
\end{equation}
Combining this bound with the general bound for all equilibrium systems, $\left( \delta c / c \right)^2 \ge 4/R_T^2$, yields the result in the main text: $\left( \delta c / c \right)^2 \ge \max(1/R_T,4/R_T^2)$.  When $R_T$ is large, systems without cooperativity perform worse than the fundamental bound by about $1/R_T$.

These arguments show that the absolute bound for equilibrium systems, $\left( \frac{\delta c} {c} \right)^2  \ge \frac{4}{R_T^2}$, can only be achieved in systems which cooperatively bind the ligand or in which multiple ligand bound to a single receptor cooperatively activate the receptor.  Without cooperativity, the bound is given by $\left( \delta c / c \right)^2 \ge \max(1/R_T,4/R_T^2)$.

\section*{Information theoretic analysis of equilibrium systems}
We consider an arbitrary equilibrium biochemical network in which receptors bind ligand and the cell uses a read-out $x$ to sense the environment.  We denote the copy numbers of the $N_s$ species in the system by the vector $\boldsymbol{N}$.  The copy numbers of $R$, $RL$, and the read-out $x$ are elements of this vector, along with any other species in the network. Since only the receptor binds the ligand, the distribution for species' copy numbers
$\boldsymbol{N}$ in the equilibrium system with $N_s$ species is given in general by (\citealp{Schmiedl2007})
 $P(\boldsymbol{N} ) = \frac{ e^{-\mu_L RL }
  \mathcal{Q}(\boldsymbol{N})}{ \Xi }$, where
$\mathcal{Q}(\boldsymbol{N}) = \prod_{i=1}^{N_s}
\frac{z_i^{N_i}}{N_i!}$ is the (canonical) partition function in terms
of the molecular partition functions $z_i$.  The grand canonical partition function, $\Xi$, normalizes the distribution by summing the numerator over all possible states consistent with the stoichiometric constraints $\mathcal{C}$: 
\begin{equation}
\Xi = \sum_{\boldsymbol{N}' \in \mathcal{C}} e^{-\mu_L RL'  } \mathcal{Q}(\boldsymbol{N'})
\end{equation}
From this distribution, it is clear that $P(x, RL, \mu_L) = P(x | RL,
\mu_L) P( RL | \mu_L) P(\mu_L) = P(x | RL ) P( RL | \mu_L) P(\mu_L)$
for any read-out $x$, so that $\mu_L \to RL \to x$ forms a Markov
chain.  That is, the chemical potential of the ligand affects the
read-out only via the instantaneous state of the receptors.  The data processing inequality then 
leads to the conclusion in the main text, $I(x; \mu_L) \le I(RL;\mu_L)$. 

The information the number of bound receptors, $RL$, has about the chemical potential of the ligand is easily bounded, since one of the few restrictions we have imposed on the equilibrium system is that the number of receptors is finite (less than $R_T$).  For any random variables $X$ and $Y$, $I(X; Y) \le H(X)$ where $H$ is the entropy of a random variable.  Furthermore, the maximum entropy distribution on a bounded support is the uniform distribution and the entropy of a discrete uniform distribution is $H(X) = \log_2(n)$ where $n$ is the number of possible states for the variable.  Thus, $I(RL; \mu_L) \le H(RL) \le \log_2(R_T+1)$.

The extensions to these
proofs when multiple types of receptors bind the ligand or when each 
receptor molecule binds multiple ligand molecules are straightforward.  Then, the quantity $R_T$ is replaced in the proofs above by the total number of ligand molecules that can be bound to receptors at any time, $L_T$.  If multiple types of receptors can bind ligand, $L_T$ is just the total number of receptors of any type.  If each receptor molecule binds more than one ligand molecule, $L_T$ is just the total number of receptors times the number of ligand molecules each receptor can bind.

\section*{Sensing error of biochemical networks viewed as discretely sampling the receptor state}
In this section we show how the sensing error of the biochemical network can be calculated by viewing the network as a discrete sampling process.  The important quantities in a sampling protocol are the number of samples taken and the spacing between them, in addition to the properties of the sampled signal.  By viewing the biochemical process as a sampling process, we mean that the underlying parameters of the biochemical network affect the sensing error only insofar as they affect these quantities, or the stochasticity in these quantities.  The benefit of viewing the network as a sampling process is that the number of samples and the spacing between them have intuitive, and well-known, effects on the sensing error: the more samples, the lower the error; the further apart the samples are, the more independent they are.  Perhaps less well known are the effects on the sensing error of stochasticity in the number of samples or the spacing between samples; these effects emerge in the process of determining the error for a discrete sampling protocol, which we do below.

We consider the biochemical networks described by the base model in
the main text and the base model plus deactivation -- the push-pull
network.  For the base model, we identified the molecules that had
collided with the receptors as samples, since these molecules' states
reflect the receptor states at the times of their collisions with the
receptor.  For the base model with deactivation, we identified the
molecules that collided with the receptor more recently than
with the phosphatase as samples.   When we refer to the number of samples, we mean the number of these molecules; when we refer to the times of the samples, we mean the times at which these molecules collided with the receptor.

We begin by rewriting the equation for the sensing error in a form that makes the connection to discrete sampling explicit, Eq. \ref{eq:noiseadd} below.  The cell senses its environment through the level of its readout $x^*$.  However, this is no different from estimating the ligand concentration from $\hat{p} = x^*/\bar{N}$:
\begin{equation}
\label{eq:fromp}
\left( \frac{\delta c}{c} \right)^2 = \frac{ \sigma_{\hat{p}}^2}{\left(\frac{d \hat{p} }{ d\mu_L} \right) ^2} = \frac{ \sigma_{\bar{x}^*}^2}{\left(\frac{d \bar{x}^* }{ d\mu_L} \right) ^2}
\end{equation}
since $\bar{N}$ is a constant, independent of $\mu_L$.  Note that the gain $d\bar{p}/d\mu_L$ is $d\bar{p}/d\mu_L  = p ( 1-p)$.  

We first consider the effect of the stochasticity in the total number
of samples, $N$.  The law of total variance allows us to decompose the
variance in the estimate $\hat{p}$ into terms arising from different
sources:
\begin{equation}
\label{eq:noiseaddv1}
\sigma_{\hat{p}}^2 =   E \left[ \text{var}(\hat{p} | N) \right] + \text{var} \left[ E  (\hat{p} |N) \right]
\end{equation}
The first term of Eq. \ref{eq:noiseaddv1} reflects the mean of the variance in $\hat{p}$ given the number of samples $N$; the second term reflects the variance of the mean of $\hat{p}$ given the number of samples $N$.

The mean and variance of $\hat{p}$ given the number of samples $N$ are more familiar quantities than their unconditioned counterparts, as we see below.  Since, by definition, the samples reflect the state of the receptor at the times $t_i$ of their collisions with the receptor, we can write the number of $x^*$ at the final time as:
\begin{equation}
x^* = \sum_{i=1}^N n_i(t_i)
\end{equation}
where $n_i(t_i)$ denotes the value of the i$^{th}$ sample --- the state of the receptor involved in the i$^{th}$  collision at the time $t_i$ of that collision, 1 if bound to ligand, 0 otherwise.  In the following, we consider a single receptor, $R_T=1$ and $n =  n_i$.  The results generalize to multiple receptors.

We can then rewrite Eq. \ref{eq:noiseaddv1}:
\begin{eqnarray}
\label{eq:noiseadd}
\sigma_{\hat{p}}^2 &=&   E \left[ \frac{N^2}{\bar{N}^2} \text{var}\left(\frac{\sum_{i=1}^N n(t_i)}{N} \middle\vert N \right) \right] \\ &+& \text{var} \left[ \frac{N}{\bar{N}}  E  \left( \frac{\sum_{i=1}^N n(t_i)}{N}  \middle\vert N \right) \right] \nonumber
\end{eqnarray}

The equation is a bit complicated, but what is important is that it
fully specifies the sensing error in terms of the number of samples,
the spacings between them, and the stochasticity in these
quantities. That is, this equation shows that the sensing error is the
error of a sampling process.  We can use it to calculate the sensing
error independently from, for example, the master equation or the linear-noise approximation.

The first term describes the error of a very standard sampling
process, one with a fixed number of samples.  We recognize the
variance
\begin{equation}
 \text{var} \left( \frac{\sum_{i=1}^N n(t_i)}{N} \middle\vert N \right) 
\end{equation}
as the error of a statistical sampling protocol in which exactly $N$ samples are taken at random times $t_i$.  This is shown explicitly in the section ``Error of discrete sampling protocols with a fixed number of samples.''  In that section, it is shown that the error for such a sampling protocol is:
\begin{equation}
 \text{var} \left(\frac{\sum_{i=1}^N n(t_i)}{N} \middle\vert N \right)  = p (1-p) \frac{1}{f_I N}
\end{equation}
where $f_I$ is the fraction of the samples that are independent, as given by Eq. 7 in the main text.
Then the first term in Eq. \ref{eq:noiseadd} is just:
\begin{eqnarray}
E \left[ \frac{N^2}{\bar{N}^2} \text{var}\left(\frac{\sum_{i=1}^N n(t_i)}{N} \middle\vert N \right) \right] = 
&E& \left[ \frac{N^2}{\bar{N}^2} p (1-p) \frac{1}{f_I N} \right]  \nonumber \\ &=&
p (1-p) \frac{1}{f_I \bar{N}}
\end{eqnarray}
That is, the first term in Eq. \ref{eq:noiseadd} is the error of a discrete sampling protocol with exactly $\bar{N}$ samples, as stated in the main text.  The only effect of the expectation in the first term is to swap $\bar{N}$ for $N$.  Dividing by the squared gain (see Eq. \ref{eq:fromp}), $d\bar{p}/d\mu_L = p(1-p)$, gives the first term in Eq. 6 in the main text.

We now turn to the second term in Eq. \ref{eq:noiseadd}.  From the law of total variance, this term describes how stochasticity in the number of samples, $N$, contributes to the sensing error.  Because the number of samples $N$ is Poisson with mean and variance equal to $\bar{N}$:
\begin{eqnarray}
\text{var} \left[ \frac{N}{\bar{N}}  E  \left( \frac{\sum_{i=1}^N n(t_i)}{N}  \middle\vert N \right) \right]  &=& 
\text{var} \left[ \frac{N}{\bar{N}}  p \right] \\ &=& \frac{p^2}{\bar{N}^2} \text{var} \left[  N  \right]  \\ &=& \frac{p^2}{\bar{N}}
\end{eqnarray}
where the probability a receptor is bound is $E[n(t_i)]=p$.  Dividing by the squared gain gives the second term in Eq. 6 in the main text.  Thus, we have derived Eq. 6 in the main text as the result of a discrete sampling protocol.

The derivations leading to Eq. \ref{eq:noiseadd} show that the sampling error for the sampling protocol must be the same as the sensing error for the biochemical network.  To check this, we can calculate the sensing error for the biochemical network, Eq. 6 in the main text, in a more standard way, determining the gain and the variance of the output $x^*$ and using Eq. 1 in the main text.  We do this for the base model with deactivation; results for the base model follow similarly.  The mean level of $x^*$ is just $\bar{x}^* = k_f p \tau_{\ell}$.  The variance in $x^*$ can be calculated using standard methods (e.g. the linear-noise approximation; see section ``Calculating the sensing error for a biochemical network''):
\begin{equation}
\sigma_{x^*}^2 = \frac{ (k_f \tau_{\ell})^2 p (1-p) }{1+\frac{\tau_{\ell}}{\tau_c}} + \bar{x}^*
\end{equation}
The gain is:
\begin{equation}
\frac{d \bar{x}^*}{d \mu_L} = p ( 1 - p) k_f \tau_{\ell}
\end{equation}
Assembling the results, Eq. 6 in the main text follows, just as it did from the sampling protocol.

\section*{The origin of the second term in Eq. 6 in the main text} The
second term in Eq. 6 in the main text emerges in the derivations above
as a consequence of the stochasticity in the number of samples $N$.
However, it is more fundamentally a consequence of the fact that the
cell does not distinguish between samples of the unbound receptor from
blank samples that do not represent a receptor state -- i.e. it does
not distinguish $x$ molecules that collided with the unbound receptor
from those that never collided with the receptor in any state.  A more
standard sampling procedure would
distinguish between these, and so would estimate $\hat{p}$ as $\hat{p}
= x^*/N$, not $\hat{p} = x^*/\bar{N}$, as above.  As we show below,
this procedure gives rise to only the first term of Eq. 6 in the main
text, allowing us to interpret the second term as the price the cell
pays for not distinguishing readout molecules that collide with the
unbound receptor from those that have never collided with the receptor
in any state.

One way to arrive at this conclusion is to imagine that all collisions
with the receptor lead to modifications of $x$. Yet, while the
ligand-bound receptor modifies $x$ into state $x^*$, the unbound
receptor modifies $x$ into another state $x^{\dagger}$. Hence, in
addition to the reaction $x + RL \to x^* + RL$ we consider the
reaction $x + R \to x^{\dagger} + R$.  Then, $N = x^* + x^{\dagger}$.
Analogously to Eq. 1 in the main text, we can then estimate the
variance of $\hat{p} = x^*/N=x^*/(x^*+x^\dagger)$ by expanding to
first order:
\begin{equation}
\delta \hat{p} \approx  g_{\hat{p},x^*} \delta x^*  + g_{\hat{p},x^{\dagger}} \delta x^{\dagger} 
\end{equation}
where the gains are:
\begin{eqnarray}
 g_{\hat{p},x^*} &=& \frac{d \hat{p}}{d x^*} = \frac{ x^*}{( x^* + x^{\dagger})^2} \\
 g_{\hat{p},x^{\dagger}} &=& \frac{d \hat{p}}{d x^{\dagger}}  = -\frac{ x^{\dagger}}{( x^* + x^{\dagger})^2}
\end{eqnarray}

The variance is then:
\begin{equation}
\sigma^2_{\hat{p}} =  g_{\hat{p},x^*} ^2 \sigma^2_{x^*} + g_{\hat{p},x^{\dagger}} ^2 \sigma^2_{x^{\dagger}} + 2 g_{\hat{p},x^*}  g_{\hat{p},x^{\dagger}}  \sigma^2_{x^*,x^{\dagger}} 
\end{equation}
where the last term accounts for the covariance.  The variances can be calculated in many ways since the system is linear.  For example, they can be calculated exactly via the linear-noise approximation.  The result is the first term of Eq. 6 in the main text, as claimed.  Indeed, there is no second term for the model described here.  This is precisely because with this scheme the number of samples $N$ is known.  While in the scheme of the main text (see Fig. 2), the system cannot discriminate between the molecules that have collided with an unbound receptor and the molecules that have not collided with the receptor at all, in this scheme the system knows exactly how many collisions there have been with the receptor: $x^*$ + $x^\dagger$.

\section*{Error of discrete sampling protocols with a fixed number of samples}

In this section, we derive the first term of Eqs. \ref{eq:noiseaddv1} and \ref{eq:noiseadd}, corresponding to Eq. 6 in the main text, as the error of a discrete sampling protocol with a fixed number of samples $N$ taken of receptor states over time.  The average receptor occupancy is estimated as:
\begin{equation}
\hat{p} = \frac{1}{N} \sum n_i(t_i)
\end{equation}
where $n_i(t_i)$ is the state of the receptor involved in the $ i^{th}$ sample at the time of that sample, 1 if the receptor was bound at time $t_i$ and  0 otherwise.  In what follows, we consider a single receptor, $R_T=1$ and $n(t_i) = n_i(t_i)$.  The results generalize to multiple receptors.  The times $t_i$ of the samples represent the times at which the molecules that store the samples of the receptor collided with the receptor.  Therefore, we choose the distribution of times between the samples to match the distribution of times between those collisions, which depends on the particular network under consideration, described below.  We count time backwards from the present time, $t=0$.  The number of samples $N$ and the distribution of times at which they were taken specifies a sampling protocol, independent of the chemical implementation.

The variance in the estimate of receptor occupancy is:
\begin{eqnarray}
\sigma_{\hat{p}}^2 &=& \text{var} \left( \frac{\sum_{i=1}^N n(t_i)}{N} \right) \\ &=& \frac{\text{var} \left( \sum_{i=1}^N n(t_i) \right)}{N^2} \\ &=&  \frac{\sigma^2}{N} + \frac{N (N-1)}{N^2} E[ \text{cov}( n(t_i), n(t_j) )  ] 
\end{eqnarray}
since $N$ is fixed, where $\sigma^2 = p (1-p)$ is the variance of the instantaneous occupancy of a single receptor.


\emph{Base model}: We first consider a statistical sampling protocol that matches the distribution of receptor-collision times of samples in the base model.  The collisions occur at random times in the interval [0,T], so we model $N$ randomly placed samples.  The time $\widetilde \Delta$ between a randomly chosen pair of uniformly distributed samples, not necessarily consecutive, is distributed as:
\begin{equation}
p(\widetilde \Delta) = \frac{2}{T} - \frac{2 \widetilde \Delta}{T^2}.
\end{equation}

Changing variables from $t_i$ and $t_j$ to $\widetilde \Delta = |t_j - t_i|$, we have $\text{cov}( n(t_i), n(t_j)  ) = \sigma^2 e^{-\widetilde \Delta/\tau_c}$.  The expectation of the covariance is then:
\begin{equation}
E[ \text{cov}( n(t_i), n(t_j) )  ]  = \sigma^2 \int e^{-\Delta/\tau_c} p(\widetilde \Delta) d \widetilde  \Delta
\end{equation}
Assembling the equations above yields the first term in Eq. 6 in the main text with $\Delta = T/N$ ($R_T=1$), where we have simplified the result with the standard assumption that $T \gg \tau_c$ and $N \gg 1$ (it does not make sense to discuss the spacing between a single sample).

\emph{Deactivation}: To take into account deactivation, we consider sampling times which match the distribution of the receptor collisions of only those $N$ molecules storing samples.  We thus have to take into account that some of the samples that have been taken are thrown away due to the deactivation process.  We begin with an alternative expression for the expected covariance:
\begin{equation}
E[ \text{cov}( n(t_i), n(t_j) )  ]  = \sigma^2 \int \int e^{-|t_j-t_i|/\tau_c} p(t_i,t_j) dt_i dt_j
\end{equation}

To match the biochemical network, the sample times $t_i,t_j$ of two samples must be independent from each other, since the collisions of different molecules with the receptor and phosphatase are uncoupled.  Therefore, $p(t_i,t_j) = p(t_i)p(t_j)$.  The marginal probability $p(t_i)$ is the probability that the collision time with the receptor of a given molecule storing a sample was $t_i$, i.e. $p(t_i | \text{sample})$.  This can be written in terms of $p(\text{sample}|t_i)$, the probability that there was a collision with the receptor at the time $t_i$ times the probability that, given a collision at that time, the associated molecule did not subsequently collide with the phosphatase: 
\begin{equation}
p(\text{sample}|t_i) dt = r dt e^{-t_i/\tau_{\ell}}
\end{equation}
Then:
\begin{equation}
p(t_i | \text{sample}) = \frac{p(\text{sample}|t_i) p(t_i)}{\int p(\text{sample}|t_i) p(t_i) dt_i } = \frac{1}{\tau_{\ell}} e^{-t_i/\tau_{\ell}}
\end{equation}   
since $p(t_i)$ is uniform.

Assembling results:
\begin{equation}
E[ \text{cov}( n(t_i), n(t_j) )  ]  = \sigma^2 \int \int e^{-|t_j-t_i|/\tau_c} \frac{e^{-t_i/\tau_{\ell}}}{\tau_{\ell}} \frac{e^{-t_j/\tau_{\ell}}}{\tau_{\ell}}  dt_i dt_j
\end{equation}
It is instructive to change variables, defining $\widetilde \Delta = |t_j - t_i|$, as before.  Then:
\begin{equation}
\label{eq:withdistr}
E[ \text{cov}( n(t_j), n(t_i) )  ]  = \int_0^{\infty} e^{-\widetilde \Delta/\tau_c} \frac{e^{-\widetilde \Delta/\tau_{\ell}}}{\tau_{\ell}} d \widetilde \Delta
\end{equation}
From this expression we can identify $p(\widetilde \Delta) = \frac{e^{-\widetilde \Delta/\tau_{\ell}}}{\tau_{\ell}}$ as the distribution of times between two randomly chosen (not necessarily consecutive) samples, when molecules can decay.  Simulations confirm this distribution.  

Completing the integral and using it in the expression for the sensing error gives the first term in Eq. 6 in the main text for the effective spacing $\Delta = 2 \tau_{\ell} / N$ (here, $R_T = 1$).  We have made the simplifying assumptions that $N \gg 1$ (it does not make sense to talk of the spacing between just one sample) and $\tau_{\ell} \gg \tau_c$, a standard assumption.  The effective spacing is not the mean nearest-neighbor spacing, but it is qualitatively similar and serves to summarize the fact that samples taken further apart in time are more independent.  Clearly, from Eq. \ref{eq:withdistr}, the error depends on the distribution of all-pairs spacings, not necessarily nearest-neighbor spacings, and it depends on the full distribution, not just the mean.

Finally, we iterate that we can perform an independent check on the derivation in this section by computing the sensing error using the linear-noise approximation, which is exact for this linear network. As mentioned, this gives exactly the same result.

\section*{No trade-offs among resources} In Fig. 3D of the main text,
we show how the sensing error depends on the pair of resources
(readout copy number $X_T$, energy $w$).  These results were obtained
via numerical minimization of Eq. 6 subject to constraints on $X_T$
and $w$.

In Fig. 3E of the main text, we show how the sensing error depends the
pair of resources (time/receptor copy number, energy).  The plot for
(time/receptor copy number, readout copy number) is the same.  In
this section, we describe the derivation of the results shown in this
figure.  In order to consider $\tau_r/\tau_c$ not necessarily large,
we need to use a form of the Berg-Purcell bound that is valid for
short integration times (\citealp{Govern2012}):
\begin{equation}
\left( \frac{\delta c}{c} \right)^2_{\rm min}  > \frac{1}{p(1-p)} \frac{1}{R_T \left(1 + \frac{\tau_r}{\tau_c} \right)}
\end{equation}
which identifies $R_T \left(1 + \frac{\tau_r}{\tau_c} \right)$ as a
limiting resource, rather than the result of the
main text, $R_T  \frac{\tau_r}{\tau_c}$, which only holds in the limit $\tau_r \gg \tau_c$.

To elucidate how the sensing error depends on (time/receptor copy
number, energy) and
(time/receptor copy number, readout copy number), we calculate the minimum sensing
error by optimizing over all parameters while fixing $R_T (1+\tau_r /
\tau_c)$ and either $w$ or $X_T$, respectively.  For a fixed $R_T
\left(1 + \frac{\tau_r}{\tau_c} \right)$ and a fixed work $w$, the
minimum sensing error is:
\begin{eqnarray}
\label{eq:fixtime}
\left( \frac{\delta c}{c} \right)^2_{\rm min} &=& \frac{w}{32} \biggl( -\frac{1}{\left( R_T \left(1+ \frac{\tau_r}{\tau_c}\right) \right)^2}  \\  &+&   \sqrt{\frac{1}{R_T \left( 1+\frac{\tau_r}{\tau_c} \right) }} \left( \frac{1}{ R_T \left(1+ \frac{\tau_r}{\tau_c} \right)} + \frac{32}{w} \right)^{3/2}  \nonumber \\ &+& \frac{128}{w^2} + \frac{80}{w \left( R_T \left(1+ \frac{\tau_r}{\tau_c} \right) \right) } \biggr) \nonumber
\end{eqnarray}
The equation for the dependence of the sensing error on (time/receptor
copy number,readout copy number) is the same, with $w$ replaced by $X_T$.  The
minimum is plotted in Fig. 3E.  The minimum tracks the worst bound,
again showing that the resources do not compensate each other.

Additional constraints on the values of rate constants will generally prevent the network from achieving these bounds.  In particular, it is common to consider that the binding of ligand to receptor is diffusion-limited, so that the bound $4/\left( R_T \left(1 + \frac{\tau_r}{\tau_c} \right) \right)$ is never achieved.  Of course, additional constraints cannot improve the performance of the network beyond the bounds required here, nor can they alter the fact that all the resources are needed for sensing.

\section*{Additional networks}

Networks are often more complicated than a simple one-level push-pull cascade.  We investigate some common motifs to understand whether they relax the trade-offs faced by sensory networks.

\emph{Multi-level cascades}: Often the signaling molecule activated by the receptor is not taken as the final read-out; rather that molecule catalyzes the activation of another molecule, and so on in a signaling cascade.  All of the molecules are reversibly degraded.  Using the same approach as for the one-level cascade, we find that the sensing error is bounded by the work done driving just the last step of the cascade: $\bar{N} \le \frac{\dot{w}_i \tau_r}{4 p}$, where $\dot{w}_i = \dot{n}_i \Delta \mu_i$  is the product of the flux of the last molecule through its cycle and the free-energy drop across that cycle, and $\tau_r$ is the slowest relaxation time in the cascade (i.e. the reciprocal of the largest eigenvalue of the relaxation matrix.)  Even more work is done at other levels of the cascade.  The results suggest that cascades do not enable more energy efficient sensing.  Additionally, each sample of an active state (bound receptor or active molecule upstream) still requires a molecule to store it.

\emph{Positive and negative feedback}: A simple model of positive feedback is autocatalysis, in which the receptor-catalyzed activation of the read-out is enhanced by the activated form of the read-out, $x^*$: $x + x^* + RL \rightleftharpoons 2 x^* + RL$.  A simple model of negative feedback can be implemented by requiring inactive $x$ for the activation: $ 2 x + RL \rightleftharpoons x + x^* + RL$.  In both cases, $x^*$ degrades according to $x^* \rightleftharpoons x$.  Neither positive feedback nor negative feedback changes the energetic requirements for sensing: $\bar{N} = \frac{\dot{n} \tau}{p} \frac{ \left( e^{ \Delta \mu_1} -1 \right) \left( e^{\Delta \mu_2} -1 \right)}{ e^{\Delta \mu - 1}}$.  As before, the free-energy drops across the reactions were calculated as the ratio of mass-action propensities.

\emph{Cooperative activation of the read-out}: If the catalytic activation of the read-out is mediated cooperatively by the receptors (i.e. $x + n RL \rightleftharpoons x^* + n RL$), then the error is reduced by a factor $n^2$ for the same amount of energy.  One way to interpret the result is that each sample requires the same amount of energy as before, but the samples are individually more informative because they reflect $n$ ligand bindings, instead of one --- indeed, the instantaneous error is lower.

\section*{Trade-offs between equilibrium and non-equilibrium sensing}

To understand how energy shapes the design of a network, we modify the push-pull network so that the read-out actually binds the ligand-bound receptor, which can boot the read-out off in a modified state: $RL + x  \rightleftharpoons RLx$, $RLx \rightleftharpoons RL + x^*$.  The active read-out decays, as before: $x^* \rightleftharpoons x$.  The reaction $RLx \rightleftharpoons RL + x^*$ coarse-grains the reactions $RLx \rightleftharpoons RLx^*$ and $RLx^* \rightleftharpoons RL + x^*$; explicitly adding these reaction gives the same results because they essentially can always be integrated out.  This network interpolates between the equilibrium and non-equilibrium networks considered in the main text.  Choosing the rate constants of the booting and decay reactions to be 0, the network reduces to the sequestration network studied in the equilibrium section.  Choosing the rate constants so that the read-out is rarely bound to the receptor, the network reduces to the push-pull network studied in the non-equilibrium section.  No resources are coarse-grained in these reductions, though the latter breaks the retroactivity of receptor-read-out binding: energy is required to break reversibility, not retroactivity.

We focus on the relationship between the number of receptors (the equilibrium resource) and the work (a non-equilibrium resource), as the network shifts from binding to catalysis.  The work is defined as $w = \dot{w} \tau_r$, as in the main text, where the relaxation time $\tau_r$ is chosen as the negative reciprocal of the smallest eigenvalue of the regression matrix of the network.  From a scaling argument and dimensional analysis, the relationship between these resources must take the form:
\begin{equation}
\frac{1}{R_{\rm T}} \left( \frac{\delta c}{c} \right)^2 \ge f \left( \frac{\dot{w} \tau_r}{R_{\rm T}} \right)
\end{equation}
for some function $f$ independent of any parameters. 

We probe this function numerically (Fig. 4).  The figure shows results from 2.5 million explicit parameter evaluations and from about 25,000 numerical minimization trials.  Minimization trials were constrained steepest descent minimizations, randomly initialized for logarithmically distributed rate constants between $\exp(-15)$ and $\exp(15)$.  To promote uniform sampling of the space, we minimized estimation error subject to constraints on the work; we minimized work subject to constraints on the estimation error; and we minimized the product of the work and the estimation error subject to constraints on either.  We also continued the best solutions over variations in the constraints to probe the global minima.  

As seen in the figure, when the work per receptor is less than about $1$ ${\rm k_BT}$, the equilibrium scheme of binding is optimal, recovering the equilibrium bound for the sensing error, $\left( \delta c/c \right)^2 \ge 4/R_{\rm T}$ (Eq. 2 in the main text with $p=1/2$).   When the work per receptor is greater than about $4$ ${\rm k_BT}$, the non-equilibrium scheme of catalysis is optimal, recovering the bound from the main text,  $\left( \delta c/c \right)^2 \ge 4/(\dot{w} \tau_r)$.  Roughly, it only makes sense to use the nonequilibrium catalysis scheme if the energy budget is sufficient to take more than one sample per receptor ($4 {\rm k_BT}$ per sample of the bound receptor), since the equilibrium scheme can take one sample of the bound receptor without any energy.  Around 1 ${\rm k_BT}$ there is an intermediate regime in which the network outperforms both these regimes by partially utilizing the bound receptor-read-out state. 

\section*{Assessing the limiting resource in biochemical networks}

In the main text, we argue that the TNF newtork could transmit much
more than one bit if it were time/receptor limited.  Here, we describe
how we arrived at that conclusion.

Even if the integration time of the network were zero and the network did
not integrate the receptor state, it would still be able to transmit
the information in the instantaneous receptor state.  The information
about the ligand concentration, $c$, in the instantaneous receptor
occupancy, $RL(T)$, is given by:
\begin{equation}
I(RL(T),c) = 1/2 \log_2 \left( \pi R_T/(2 e) \right)
\end{equation}
To arrive at this result, we calculated the information transfer of a
biochemical system that takes the receptor occupancy, and not a
downstream readout, as the final output.  We assumed simple
ligand-binding kinetics, $R+L \leftrightharpoons RL$, and assumed that
ligand binding is not affected by any downstream processes.  More
complicated kinetics (e.g. cooperativity) would likely increase the
instantaneous information transfer.  The result assumes that the
ligand-binding kinetics are optimized with respect to the distribution
of input concentrations of the ligand; i.e. the information transfer
calculated is the channel capacity of the network.  The channel
capacity is the appropriate quantity to consider, because it is the
experimentally reported quantity in the paper by
(\citealp{Cheong:2011fk}).  We followed the method in
\cite{Tkacik:2008kx} to calculate the channel capacity.

TNF signaling utilizes $R_T=2000$ receptors on the cell surface
(\citealp{Guo:1999bh}), corresponding to $1/2 \log_2 \left( \pi R_T/(2
  e) \right) = 5$ bits of information.  If the network integrates the
receptor state, the information could be even higher.  The fact that
the actual information transfer is instead much less than 5 bits
suggests that receptors/time do not limit the accuracy of sensing, but
rather another resource, such as copy numbers of signaling components
or energy.

The following paragraphs address various nuances to the above
argument.  First, note that restrictions on the probability
distribution of inputs can prevent the system from achieving the
channel capacity.  This is true both for our bound and for the
calculated information transmission through the entire network in
the paper by Cheong {\it et al.} One biologically relevant restriction
on the probability distribution of inputs is the support of the
distribution, particularly the maximum biologically relevant
concentration of the ligand; if achieving the channel capacity
requires input distributions with large probability for concentrations
that are much higher than those biologically observed, then the
channel capacity is not really a relevant measure for the
capacity of the network.  Important in this context is that the
dissociation constant $K_D$ for TNF binding is 0.323 ng/mL
(\citealp{grell1998}), about the same as the half-saturation for the
TNF response as measured by Cheong {\it et al.}.  So achieving the
channel capacity at the level of the receptors does not require higher
concentrations than achieving the channel capacity of the whole
network.  This means that, while restrictions on the maximum input
concentration would prevent the system from achieving the channel
capacity of 5.5 bits at the level of the receptors, they would also
prevent the system from achieving the channel capacity of 1 bit at the
level of the output, maintaining the discrepancy.

The above arguments assume 
  that a) the principal role of the signaling network is to time
  integrate the receptor, and b) that this improves information
  transmission if energy and the copy numbers of the signaling
  molecules are not limiting, and the network is hence not too
  noisy. However, signaling systems with enough fuel and signaling
  molecules that time-integrate the receptor, do not necessarily
  increase information transmission. They can also reduce information
  transmission by collapsing many input states onto the same output
  state. This can happen when the input-output relation is (strongly)
  non-linear.  However, the experimental data in the paper by Cheong
{\it et al.} suggest that this is not the case for the TNF
  network, as the response increases mono-modally and gradually with
  the input. In fact, the output is also noisy. Indeed, the authors
  attribute the loss of information transmission to biochemical
  noise, which, according to our analysis, could be due to
    limiting amounts of readout molecules or energy.

A final note is that while the above arguments show that the number of
receptors and time are not limiting and suggest that downstream
molecules or energy are limiting, it is cannot be ruled out that other sources
of noise, which we have not modeled, are instead limiting. For example, the sensing precision could
  be limited by cell-to-cell variability in the copy numbers of
  signaling molecules (expression or capacity noise
(\citealp{Colman-Lerner:2005dq})).  These could even involve
  variations in the number of receptors themselves. However,
  back-of-the-envelop calculations suggest that such variations are
  not enough to explain the discrepancy above.  Moreover, many
  biological systems, including some two-component systems, are
  insulated against fluctuations in protein expression (\citealp{Kollmann:2005fz}),
  supporting the idea that in these cases energy or protein copy numbers are indeed limiting the accuracy of
    sensing.

\bibliographystyle{prsty}

\end{document}